\documentclass[12pt,USenglish]{article}
\pdfoutput=1

\usepackage{setspace}
\usepackage{cite}

\renewcommand{\Phi}{\phi}

\usepackage[utf8]{inputenc}
\usepackage{geometry}
\usepackage[USenglish]{babel}
\usepackage{verbatim}
\usepackage{prettyref}
\usepackage{amsmath}
\usepackage[normalem]{ulem}
\usepackage{amssymb}
\usepackage{graphicx}
\usepackage{setspace}
\usepackage{esint}

\usepackage{color}
\definecolor{darkgreen}{rgb}{0,0.5,0}
\definecolor{darkblue}{rgb}{0,0,0.6}
\definecolor{purple}{rgb}{0.4,.2,0.7}
\definecolor{orange}{rgb}{0.95, 0.5, 0.3}

\usepackage[bookmarks=false,colorlinks=true,citecolor=darkgreen,linkcolor=darkblue,urlcolor=blue]{hyperref}

\numberwithin{equation}{section}
\numberwithin{figure}{section}
\numberwithin{table}{section}

\def\be{\begin{equation}}
\def\ee{\end{equation}}
\def\bea{\begin{eqnarray}}
\def\eea{\end{eqnarray}}
\def\ba{\begin{align}}
\def\ea{\end{align}}

\def\cO{{\cal O}}

\begin{document}
\begin{spacing}{1.3}

~
\vskip5mm

\begin{center} {\Large \bf Eigenstate thermalization\\ in the Sachdev-Ye-Kitaev model}

\vskip10mm
Julian Sonner \& Manuel Vielma\\
\vskip1em
{\it  Department of Theoretical Physics, University of Geneva\\ 24 quai Ernest-Ansermet, 1214 Gen\`eve 4, Switzerland} 
\vskip5mm

\tt{ \{julian.sonner, manuel.vielma\}@unige.ch\,}

\end{center}

\vskip10mm

\begin{abstract}
The eigenstate thermalization hypothesis (ETH) explains how closed unitary quantum systems can exhibit thermal behavior in pure states. In this work we examine a recently proposed microscopic model of a black hole in AdS$_2$, the so-called Sachdev-Ye-Kitaev (SYK) model. We show that this model satisfies the eigenstate thermalization hypothesis by solving the system in exact diagonalization. Using these results we also study the behavior, in eigenstates, of various measures of thermalization and scrambling of information. We establish that two-point functions in finite-energy eigenstates approximate closely their thermal counterparts and that information is scrambled in individual eigenstates. We study both the eigenstates of a single random realization of the model, as well as the model obtained after averaging of the random disordered couplings.  We use our results to comment on the implications for thermal states of a putative dual theory, i.e. the AdS$_2$ black hole.
\end{abstract}

\pagebreak 
\pagestyle{plain}

\setcounter{tocdepth}{2}
{}
\vfill
\tableofcontents
\section{Introduction}
Via holographic duality (or `AdS/CFT') black holes are described by thermal states of a dual quantum field theory. The process of black hole formation and evaporation  corresponds to the process of thermalization of certain (unitary) quantum field theories, evolving from non-equilibrium initial states towards thermal equilibrium. However, many important questions pertaining to this attractively simple picture remain mysterious, as expressed notably in the information loss paradox \cite{hawking1976breakdown}, and its more recent ramifications \cite{Almheiri:2012rt}. An important aspect of a satisfactory resolution of these puzzles within holography is a precise understanding of the relation between information loss, unitary evolution and thermalization of the boundary theory. Put succinctly the question of how thermal  \cite{deutsch1991quantum,srednicki1994chaos}  a generic high-energy eigenstate looks in a theory with a holographic dual is closely tied to the question of how legitimately one may consider the dual geometry of this pure state to be a black hole.  It is therefore important to understand the detailed mechanism of thermalization in any given dual field theory model, as well as to extract lessons for the general class of theories with bulk duals, wherever these are available. As is often the case, the most promising starting points are the simplest instances which still contain enough structure to allow one to address the subtleties also present in the higher-dimensional, more complicated theories of primary concern. For this reason, lower-dimensional holographic models, such as AdS$_3$/CFT$_2$  (i.e. three-dimensional gravity) \cite{Fitzpatrick:2015foa,Fitzpatrick:2016ive,Anous:2016kss,Anous:2017tza}, as well as the even lower-dimensional case of NAdS$_2$/NCFT$_2$ (`Near AdS$_2$', `Near CFT$_2$') \cite{AK15,Sachdev:2015efa,Maldacena:2016hyu,Maldacena:2016upp} have once again risen to the fore.

The subject of the present study is a recently proposed many-body quantum mechanical system of interacting fermions \cite{AK15,Sachdev:2015efa}, closely related to an older one with a controllable spin-glass ground state \cite{Sachdev:1992fk}, the so-called Sachev-Ye-Kitaev (SYK) model. This is a quenched disorder quantum system with random all-to-all couplings, which shows emergent approximate conformal invariance in the infrared as well as an extensive entropy at low temperature \cite{AK15,Sachdev:2015efa,Polchinski:2016xgd,Maldacena:2016hyu}.  Furthermore, it has also been demonstrated \cite{AK15} that the model exhibits maximal quantum chaos (in the sense of \cite{Maldacena:2015waa}), diagnosed by a certain Lyapunov exponent ($\lambda_L$) extracted from thermal four-point correlation functions \cite{AK15,Polchinski:2016xgd,Maldacena:2016hyu} (see also \cite{Bagrets:2016pxi,Bagrets:2017pwq} for interesting aspects of these correlation functions).  In the SYK model then, the question of thermalization is one about the behavior of a simple\footnote{Simple enough that various experimental approaches have been put forward \cite{Danshita:2016xbo,Garcia-Alvarez:2016wem,Pikulin:2017mhj,Chew:2017xuo}.} many-body Hamiltonian under unitary evolution, i.e. the study of thermalization in closed many-body  quantum systems. Much has  been learned about such situations (and we refer the reader to the review \cite{d2015quantum} for more details and references). The simplicity of the SYK model means that it is a perfect candidate for a detailed study of the mechanism of thermalization within the setting of holographic duality with access to the deeply quantum regime.

It should be kept in mind that owing to the ensemble average implicit in the model, there are potential subtleties in connecting results in the  SYK model to black-hole physics. To start addressing this issue, \cite{Witten:2016iux,Klebanov:2016xxf,Gurau:2016lzk} have shown that certain tensor theories defined without averaging over an ensemble, give rise to the same large-$N$ limit as the SYK models. Here we also study the question of thermalization in a given random realization of the SYK model, for which the previous cautionary remark does not apply. To the extent that one limits oneself to self-averaging quantities the same then holds for the disorder averaged model.

In this paper we demonstrate in detail that the SYK model satisfies the properties of the eigenstate thermalization hypothesis and therefore establish this scenario as the appropriate mechanism for thermalization in this prototypical example of holographic duality. We expect that other versions of holographic duality also exhibit eigenstate thermalization, and we offer some further comments in the discussion section. A discussion on ETH and holography may be found in \cite{Khlebnikov:2013yia}. We are here studying a version of the SYK model in which four fermions interact, but one can extend this to a $q-$fermion interaction \cite{Maldacena:2016hyu}. This was used by \cite{Eberlein:2017wah} in order to analytically study thermalization following a coupling quench. Our study elucidates the thermal structure inherent in pure states more generally, and it would be enlightening to derive some of our numerical results analytically, perhaps starting at large $q$. The free model $q=2$ was already shown to satisfy eigenstate thermalization in \cite{Magan:2015yoa}.
\subsection{Summary of results}
Let us briefly summarize the main results of this paper.  First and foremost we establish, by exactly diagonalizing the complex SYK Hamiltonian for up to $17$ sites, that expectation values of non-extensive -- that is those involving a few sites only -- operators are to a very good approximation thermal. In particular their matrix elements take on the expected form encapsulated in the eigenstate thermalization hypothesis. This has interesting ramifications for the holographic dual, and we address some of these in the discussion section.

By studying off-diagonal matrix elements of non-extensive operators we establish that the SYK model behaves like a random-matrix theory (RMT) for a certain range of energies, but more generally deviates from such RMT behavior. By analogy with the theory of disordered conductors we refer to the energy scale at which deviations from RMT are observed as the Thouless energy $E_T$. We find that the Thouless energy is controlled by the coupling as $E_T \sim J^2$. The more strongly coupled the system, the larger the energy range for which it exhibits RMT behavior.

Having established that eigenstates behave thermally we compare correlation functions of non-extensive operators in eigenstates with their corresponding thermal expectation values. We find that two-point and four-point correlation functions in eigenstates are qualitatively similar to thermal averages, but do differ in their detailed structure. We also study correlations in large superpositions of eigenstates which approximate thermal averages with respect to a canonical density matrix. To the extent that we can extrapolate these results to large values of $N$ this suggests that individual eigenstates can in some sense be considered to be dual to the black-hole geometry in the putative dual, although we make no statement about its interior (see discussion section).

This motivates us to consider measures of scrambling in eigenstates, which we find to behave in accordance with expectations from combining known results in the canonical ensemble with our results on eigenstate thermalization. We therefore expect that there is a large$-N$ eigenstate equivalent of the maximal scrambling exponent satisfied by the SYK model, as detailed in (\ref{eq.EstateLyapunov}) below.

\section{Background}
In this section we introduce the model to be examined, discuss its properties, and introduce some pertinent notions of many-body thermalization necessary to follow the remainder of the work. We begin by preparing the ground for our investigation of eigenstate thermalization, followed by a definition of the model along the lines of \cite{AK15,Sachdev:2015efa}. Much of the material in this section is well known, and we refer the reader to the review \cite{d2015quantum} for further details. We nevertheless include some background material in an effort to make the paper more self contained.

The manifestation of thermal behavior, defined as the applicability of equilibrium thermodynamics to isolated quantum systems at first seems puzzling from a microscopic perspective: how is a seemingly mixed thermal state reached, starting from a pure initial quantum state? A common test case is the so-called quench scenario\footnote{See \cite{Eberlein:2017wah} for a recent explicit study of quenches in SYK as well as \cite{Kourkoulou:2017zaj}. In the semiclassical regime of the gravity dual this corresponds to thermalization in AdS$_2$, as studied recently e.g. in \cite{Erdmenger:2016msd,Withers:2016lft}. }, where a system is prepared in a far-from-equilibrium state and its thermalization is studied as the system evolves in time. In classical mechanics the emergence of thermal behavior as the end point of such an evolution is a consequence of dynamical chaos, or more formally of ergodicity: the delocalization of a general initial state over phase space is understood as a consequence of dynamical chaos. The end-point is then given as the state of maximal entropy, consistent with the values of its (few) integrals of motion at the initial time.
In quantum systems, where the dynamics is linear and therefore not chaotic in the classical sense, the detailed mechanism is different, even though the notion of (quantum) chaos again plays a central role. A powerful framework for quantum thermalization is given by the eigenstate thermalization hypothesis \cite{deutsch1991quantum,srednicki1994chaos,rigol2007thermalization} (ETH). In this scenario a quantum system thermalizes as a consequence of the structure of its eigenstates, rather than chaotic dynamics. In effect thermal behavior is already apparent at the level of individual eigenstates of the many-body Hamiltonian. 
\subsection{Eigenstate thermalization \label{sec.ETH}}
The Eigenstate Thermalization Hypothesis \cite{deutsch1991quantum,srednicki1994chaos} is an assumption on the properties of the spectrum of many-body Hamiltonians for systems showing thermal behavior. We start by stating the ETH ansatz for matrix elements of  nonextensive operators\footnote{One often sees ETH stated for local operators. In the present context this is not an appropriate choice of words, since the model itself is fully nonlocal. What we mean instead are operators that involve only a finite number of lattice sites that does not scale with $N$. Clearly local operators in a local theory satisfy this property.} and then proceed to a discussion of its motivation and some consequences. For a system satisfying ETH the matrix elements of local operators (or non-extensive operators, as is the case for us), evaluated in the eigenbasis of the Hamiltonian take the form
\be\label{eq.ETHHypothesis}
\langle m | \hat{\cal O} | n \rangle = \overline{{\cal O}(\bar E)} \delta_{mn} + e^{-S(\bar E)/2} f_{\cal O}(\bar E,\omega) R_{mn}\,,
\ee
referred to as the ETH ansatz. Here $\overline{{\cal O}(\bar E)}$ is a smooth function of the average energy $\bar E = \frac{E_m + E_n}{2}$ and $f_{\cal O}(\bar E, \omega)$ is a smooth function of the difference, $\omega = E_m-E_n$, in addition to $\bar E$. The remainder function $R_{mn}$ is a Gaussian random variable (real or complex) with zero mean and unit variance.
The most striking feature of this ansatz is that off-diagonal matrix elements are suppressed by a large number, more precisely the thermodynamic entropy $S(\bar E)$, while diagonal elements are of order one.

The significance of ETH is that it explains thermal behavior of certain operators in closed quantum systems via properties of stationary states of the Hamiltonian. Thermalization for local (or non-extensive) operators occurs in this picture in the sense that long-time averages of observables
\be
\overline{\cal O} :=\lim_{T\rightarrow\infty}\frac{1}{T}\int^T_0  \left\langle \psi| {\cal O}(t) |\psi \right\rangle dt
\ee
approach the diagonal ensemble
\be
\overline{\cal O} = \sum_{n=1}^{{\rm dim}({\cal H})}|c_{n}|^2{\cal O}_{nn}\,,
\ee
where the $c_n$ are the expansion coefficients of the initial state in terms of energy eigenstates, $\left.|\psi\right\rangle = \sum_n c_n | n\rangle\,$.
 The diagonal ensemble in turn becomes equivalent to the microcanonical ensemble if the matrix elements ${\cal O}_{nn}$ are continuously varying functions of the average energy $\overline{E}$, as required in (\ref{eq.ETHHypothesis}). Thus, by ensemble independence, thermodynamic expectation values are recovered on average at late time.  The picture is thus that dephasing essentially reduces an initial state to the diagonal ensemble and that the remaining diagonal values are themselves thermal with fluctuations suppressed by the size of the Hilbert space. In this way ETH is a satisfactory scenario for the emergence of thermal behavior in closed quantum systems under unitary evolution. In this work we establish eigenstate thermalization for the complex spinless Fermion version of the SYK model. We now describe the basics of this system.

\subsection{Complex SYK model \label{sec.SYKModel}}
The model involves fermionic microscopic degrees of freedom on $N$ sites. The spatial configuration of these sites is immaterial, due to the non-local nature of the interaction. In this paper we work with complex spinless fermions subject to the  Hamiltonian \cite{Sachdev:2015efa,Fu:2016yrv}
\be\label{eq.SYKHam}
H = \sum_{i,j,k,l}^NJ_{ij;kl}c^\dagger_i c^\dagger_j c_k c_l\,
\ee
with complex coupling parameters $J_{ij;kl}$.
We define a set of fermion creation and annilihation operators, satisfying
\be
\{ c_i, c_j^\dagger\} = \delta_{ij}\,,\qquad \{c_i, c_j\} = \{ c_i^\dagger, c_j^\dagger\} = 0\,.
\ee
Antisymmetry of the fermions  as well as Hermiticity of the Hamiltonian impose the constraints
\be
J_{ij;kl} = -J_{ji;kl}\,,\quad J_{ij;kl} =- J_{ij;lk}\,,\quad J_{ij;kl} = J_{kl;ij}^* \,.
\ee
The remaining independent coupling constants in $J_{ij;kl}$ are drawn from a Gaussian distribution with zero mean. The variance
\be
\overline{\left|J_{ij;kl}\right|^2}=\frac{3!J^2}{N^3}\,
\ee
sets the average strength of the coupling, which has to scale with the number of sites $N$ as shown to ensure a well-behaved large-$N$ limit. We will keep this normalization despite the fact that we always work at strictly finite $N$. The Hamiltonian, as written above (\ref{eq.SYKHam}) does not respect particle-hole symmetry. However, the particle-hole violating effects come from terms where two indices of $J_{ij;kl}$ take the same value and so they are suppressed by powers of $1/N$. One could add a chemical potential that restores particle hole symmetry at any $N$, as in \cite{Fu:2016yrv}. Since both versions connect to the same large-$N$ limit we have chosen here instead to work with the simpler Hamiltonian  (\ref{eq.SYKHam}).

\subsubsection{Properties}

The Hamiltionian (\ref{eq.SYKHam}) is a variant of a set of models considered previously in the context of quantum spin glasses \cite{Sachdev:1992fk}, but crucially does not itself have a spin-glass phase. A closely related model has recently been proposed by Kitaev in terms of Majorana fermions at $N$ sites  \cite{AK15}. 

The model is solvable\footnote{So far, expressions have been obtained for the two-point, four-point and six-point functions \cite{Sachdev:1992fk,parcollet1999non,Sachdev:2015efa,Polchinski:2016xgd,Jevicki:2016bwu,Gross:2017hcz,Jevicki:2016ito,Dartois:2017xoe}.} at large-$N$ and can be shown to exhibit a number of striking properties \cite{AK15,Polchinski:2016xgd,Maldacena:2016hyu}, chiefly among them the fact that it exhibits maximal chaos, in the sense that an appropriately defined (see section \ref{sec.OTOC}) out of time order four-point function (OTOC) decays exponentially for times up to the scrambling time,
\be
\langle A(t)B(0) A(t) B(0) \rangle_\beta \sim 1-\alpha e^{\lambda_L t}\,,
\ee
where $\alpha$ is a coefficient, e.g. $\alpha \sim \beta J/N$ for the large-$N$ limit of SYK \cite{Maldacena:2016hyu}. This defines a quantum version of a Lyapunov exponent. The average is taken in the thermal state, as indicated. This Lyapunov exponent takes the maximal \cite{Maldacena:2015waa} value $2\pi/\beta$ in the SYK model, which is the same as that of a Schwarzschild black hole in Einstein gravity \cite{Shenker:2013pqa,AK15}. Here we provide numerical evidence that this Lyapunov exponent can also be extracted by considering instead energy eigenstates, that is to say correlation functions of the form
\be\label{eq.estateLyapunov}
\langle E| A(t)B(0) A(t) B(0) |E\rangle\sim 1-\alpha e^{\lambda_L t}\,.
\ee
The value of $\lambda_L$ in eigenstates can be meaningfully compared with the thermal value by appealing to the eigenstate thermalization hypothesis in order to associate a temperature $T = \beta^{-1}$ to the individual eigenstate $|E\rangle$. To this end one may either follow \cite{rigol2009quantum} and associate an effective temperature $\beta^{-1}$ with energy $E$ via the canonical average
\be\label{eq.canonicalEnergyTemperatureMap}
E(\beta)  = \frac{1}{Z}{\rm Tr}\left[e^{-\beta H}H \right]\,.
\ee
Alternatively, for many of our computations, we determine all thermodynamic quantities in a microcanonical ensemble centered at the average energy $\overline{E} = (E_n + E_m)/2$  for ease of comparison with the ETH ansatz.

Here we study (\ref{eq.estateLyapunov}) using exact diagonalization.  It would clearly be interesting to establish some of our results analytically at large $N$, or perhaps via the large-D limit of \cite{Ferrari:2017ryl}. Our work can be viewed as motivating the conjecture that $\lambda_L$, defined in terms of eigenstates as in (\ref{eq.estateLyapunov}) will satisfy
\be\label{eq.EstateLyapunov}
\lambda_L = 2\pi / \beta(\bar E)
\ee
when evaluated at large $N\gg 1$ and large coupling $\beta J \gg 1$, where $\beta$ is defined as the inverse of the map (\ref{eq.canonicalEnergyTemperatureMap}) above.
Let us now move on to a discussion of the methods and results of this work.

\section{One point functions and eigenstate thermalization\label{sec.specTherm}}

The main analysis technique in this work is numerical diagonalization of the many-body Hamiltonian.
For most applications we numerically diagonalize the Hamiltonian (\ref{eq.SYKHam}) for up to $17$ sites\footnote{We emphasize that this corresponds to a Hilbert space dimension $2^N=2^{17}$ which is the same as that of the Majorana SYK model with $M=34$ sites, corresponding to a Hilbert space dimension of $2^{M/2}$. }. This allows us to explicitly calculate the matrix elements of non-extensive operators made up of creation and annihilation operators involving a small number of sites. 

It is easily seen that total fermion number
\be
\hat n_{\rm F} = \sum_{i}^N c_i^\dagger c_i
\ee
commutes with the Hamiltonian
\be
\left[H,\hat n_{\rm F}  \right] = 0\,,
\ee
allowing us to work in sectors of fixed fermion number $n_{\rm F}$, denoting the filling fraction $\nu = \frac{n_F}{N}$. This is useful numerically as it allows us to cut down the effective matrix sizes in the actual diagonalization process. For the most part we will work in the half-filling sector $\nu = \frac{1}{2}$. If the number of sites $N$ is odd, we mean $\nu = \frac{N+1}{2N}$ when we refer to `half filling'.

\subsection{On-diagonal terms are thermal\label{sec.OnDiag}}
\begin{figure}[t!]
\begin{center}
\includegraphics[width=0.49\textwidth]{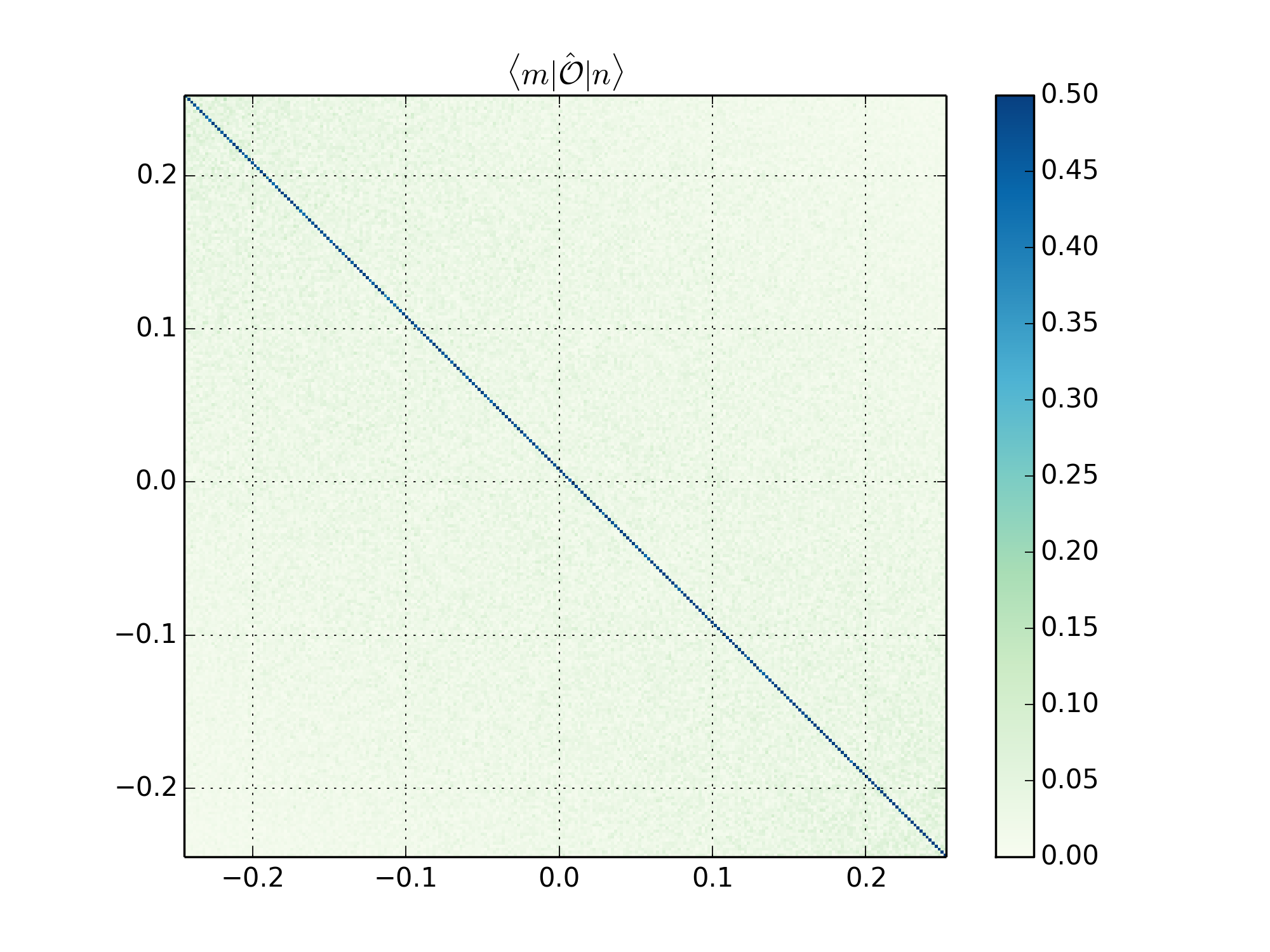} \includegraphics[width=0.49\textwidth]{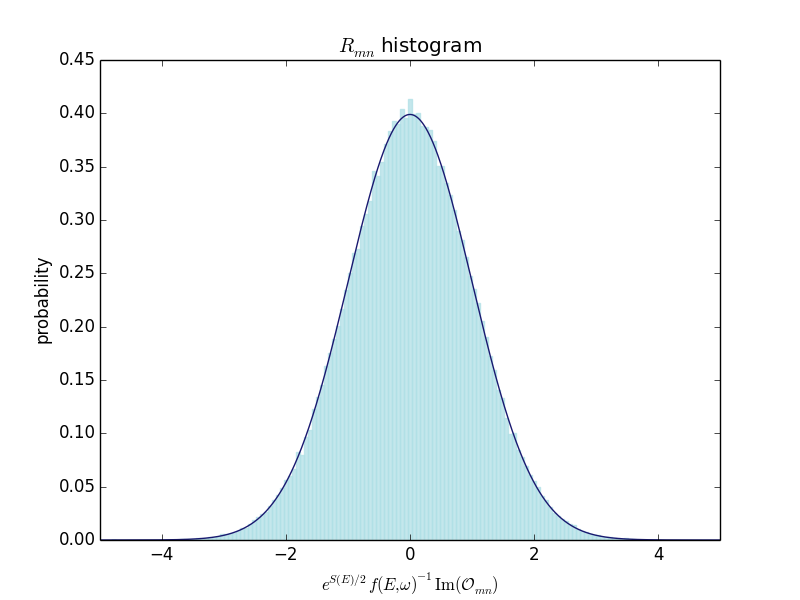}
\caption{\small Absolute values of matrix elements $|{\cal O}_{nm}| = \left|\langle n | {\cal O} | m \rangle\right| $  for the single-site number operator ${\cal O} = \hat n_N$ at half filling $\nu = \frac{1}{2}$.  {\bf Left panel}: we show the absolute values of matrix elements against their energies $E_n/J$ labelled along horizontal and vertical axes for a single realization at $ N=10$. We have checked this behavior for higher values of $N$, and found excellent agreement with ETH expectations. {\bf Right panel: } Histogram of the remainders $R_{mn}$ for $1000$ realizations at $N=12$. As we see these are accurately fit by a unit width Gaussian with zero mean confirming the ETH ansatz (\ref{eq.ETHHypothesis}). Again we have verified this for other accessible values of $N$. Similar results are obtained for models with short-range interactions in \cite{beugeling2015off}. \label{fig:SYKnMatrixl}}
\end{center}
\end{figure}

\begin{figure}[t!]
\begin{center}
\includegraphics[width=0.48\textwidth]{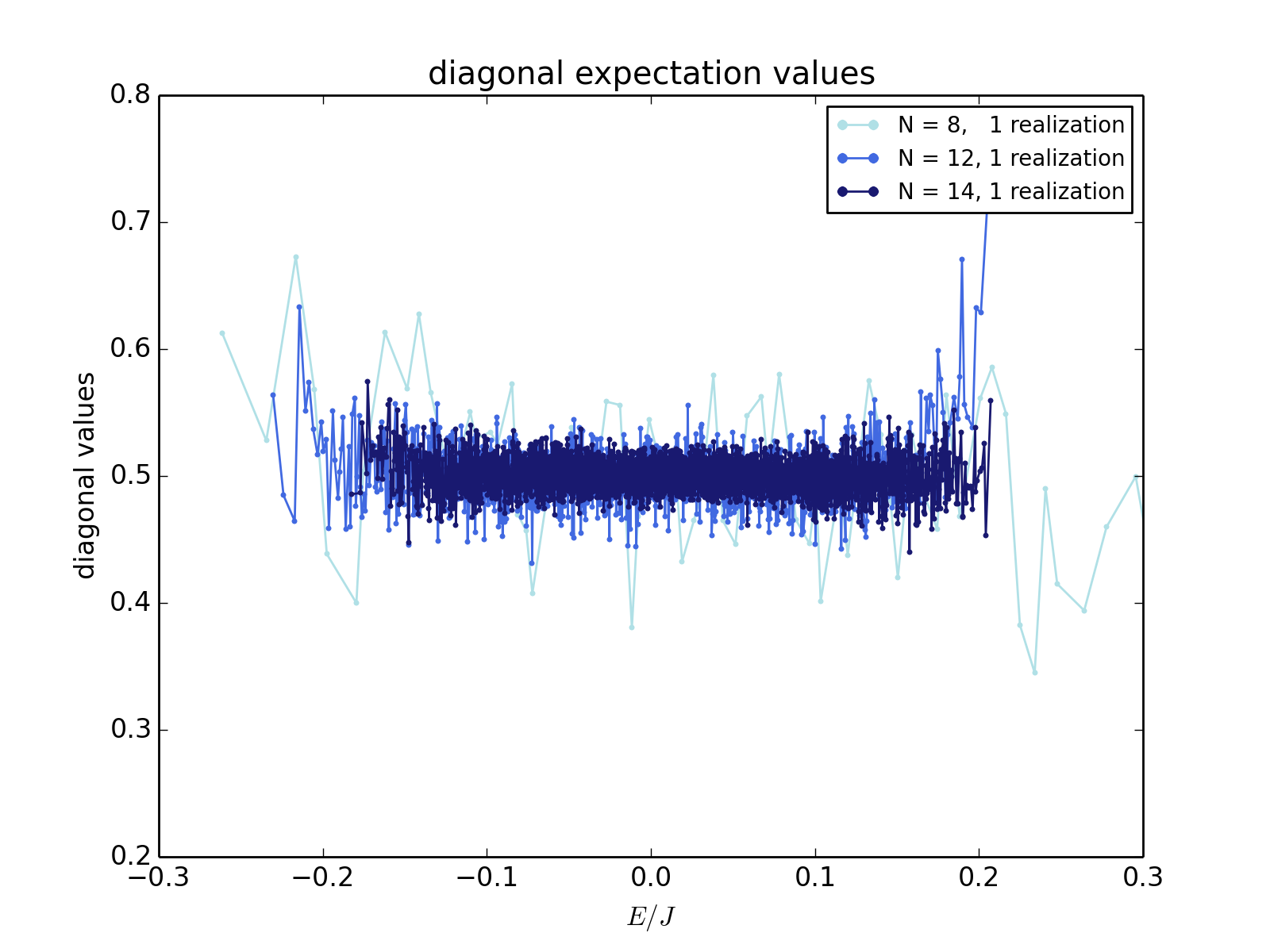}\, \includegraphics[width=0.48\textwidth]{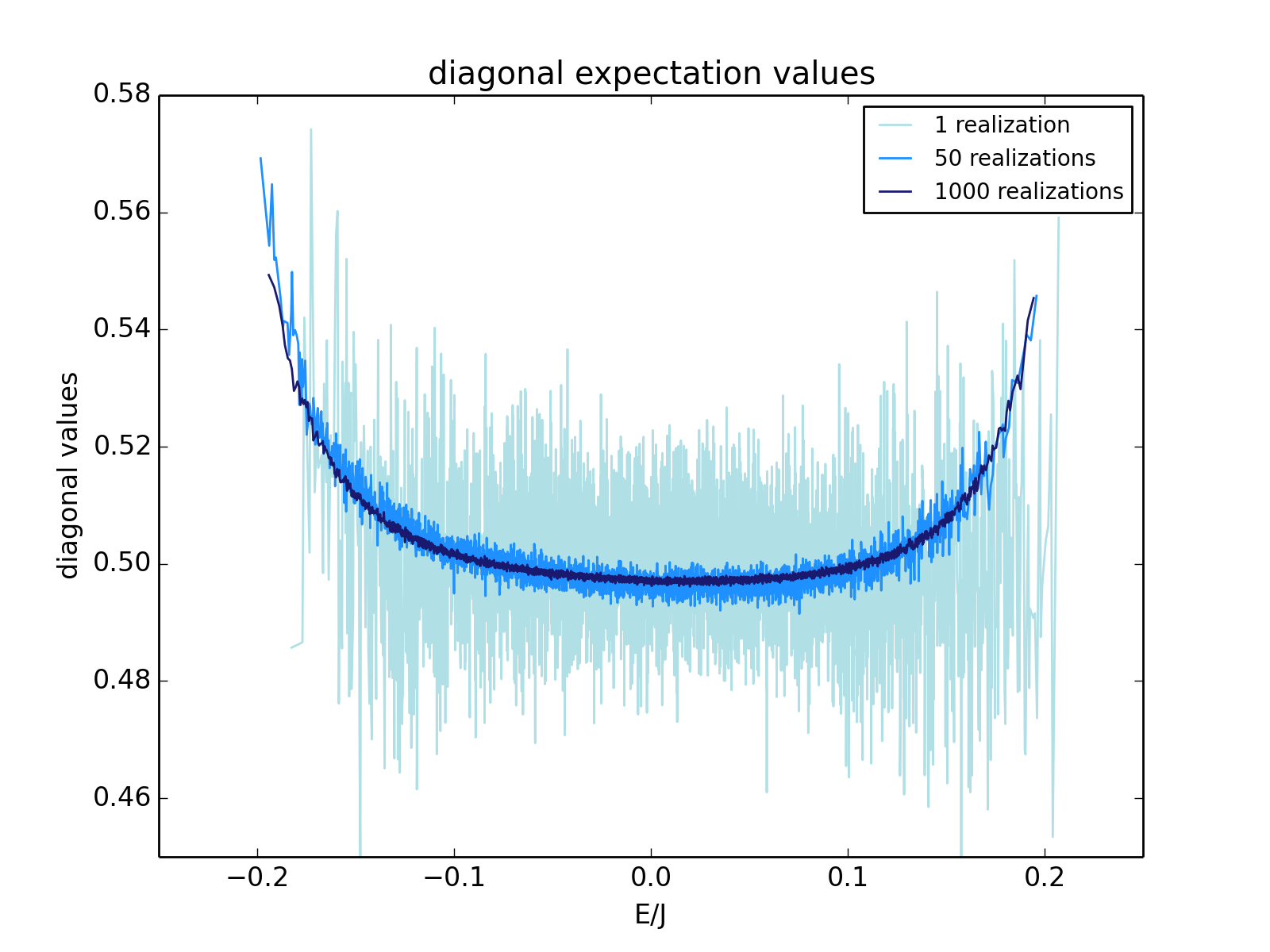}\,\includegraphics[width=0.48\textwidth]{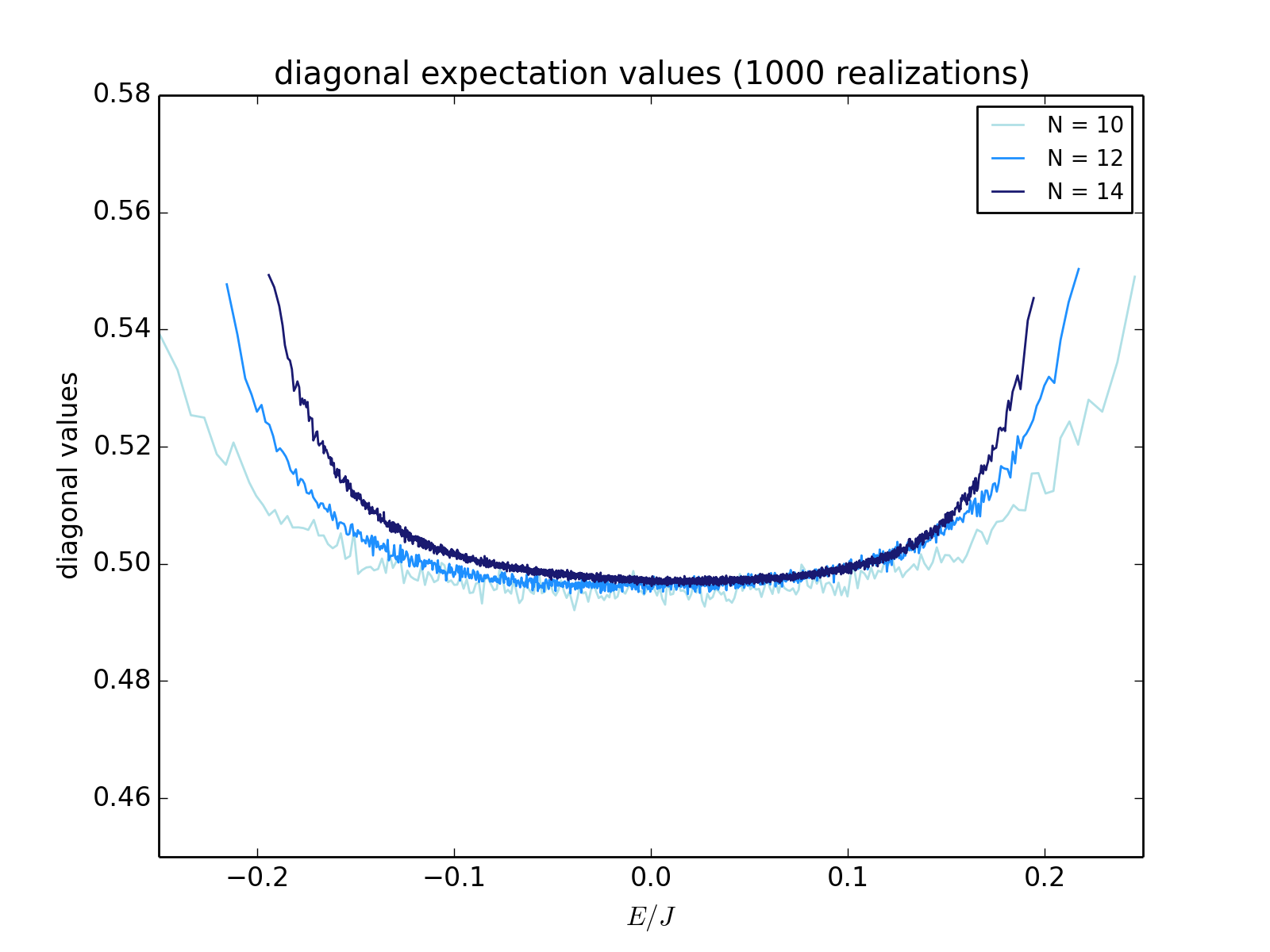}
\caption{\small Diagonal expectation values for the single-site number operator at site $N$,  that is $\hat n_N$ at half filling $\nu = \frac{1}{2}$.  {\bf Top panel}:  we show a single random realization for increasing Hilbert space dimension corresponding to $N=8, 12, 14$. We see that the on-diagonal expectation values of a single realization approach closer and closer to a smooth curve.  {\bf Left panel}:  we show the effect of averaging of the random couplings at given fixed Hilbert space dimension, $N=14$. As expected the on-diagonal values of the ensemble approach closer and closer to a smooth curve.  {\bf Right panel}:  we show the limiting curves for the model with fixed Hilbert space dimension corresponding to $N=10,12,14$, averaged over $1000$ realizations. \label{fig:SYKnNdiagonal}}
\end{center}
\end{figure}

According to the ETH ansatz (\ref{eq.ETHHypothesis}), diagonal matrix elements $ {\cal O}_{nn}=\langle n | {\cal O} | n\rangle$ are smooth functions of the average energy $\bar E$, while off-diagonal elements are suppressed by the entropic factor $e^{-S/2}$.

We start by illustrating this exponential suppression in Figure (\ref{fig:SYKnMatrixl}), where it is easily seen that only the diagonal entries of the matrix are appreciable. We illustrate this behavior for $N=10$ in Figure (\ref{fig:SYKnMatrixl}), but have checked it extensively for other accessible values of $N$ finding excellent agreement with ETH expectations. For $N=10$, one can make out the fluctuating nature of the off-diagonal matrix elements, which we will characterise precisely in section \ref{sec.OffDiag} below. Of course so far this is at the level of qualitative observation and we will turn to a more quantitive analysis of the off-diagonal matrix elements below.

 Before we do so, let us analyse the on-diagonal matrix elements, ${\cal O}_{nn}\left(\bar E\right)$ in detail. The expectation for finite values of $N$ is that the diagonal matrix elements condense more and more tightly onto a limiting smooth curve $\overline{{\cal O}(\bar E)}$ which is defined by extrapolation to the thermodynamic limit $N\rightarrow \infty$. This is illustrated in Figure \ref{fig:SYKnNdiagonal}. As mentioned before, for a model that involves a disorder average, such as SYK, we should distinguish between what happens in an individual realization, and what happens in the ensemble. The convergence towards a limiting curve for a single realization is shown in the top panel of Figure \ref{fig:SYKnNdiagonal}, while the convergence due to disorder averaging is shown in the bottom two panels of that same Figure. If a certain property is satisfied in both senses, i.e. for a single realization as well as in the disorder averaged theory, this property is said to be self-averaging. Here we confirm that the diagonal part of the ETH ansatz in the SYK model is satisfied both in a single realization and in the disordered theory. Of course this is only true for sufficiently large Hilbert space dimension. 
 
  We have further verified this property for a number of different non-extensive operators over a  range of filling fractions. We show a representative selection of these results for the hopping operator for two fixed sites $h_{ik} = c^\dagger_i c_k + c^\dagger_k c_i$ for different values of $N$ in Appendix \ref{app.hopping}. 

\subsection{Off-diagonal terms \label{sec.OffDiag}}
\begin{figure}[t!]
\begin{center}
\includegraphics[width=0.48\textwidth]{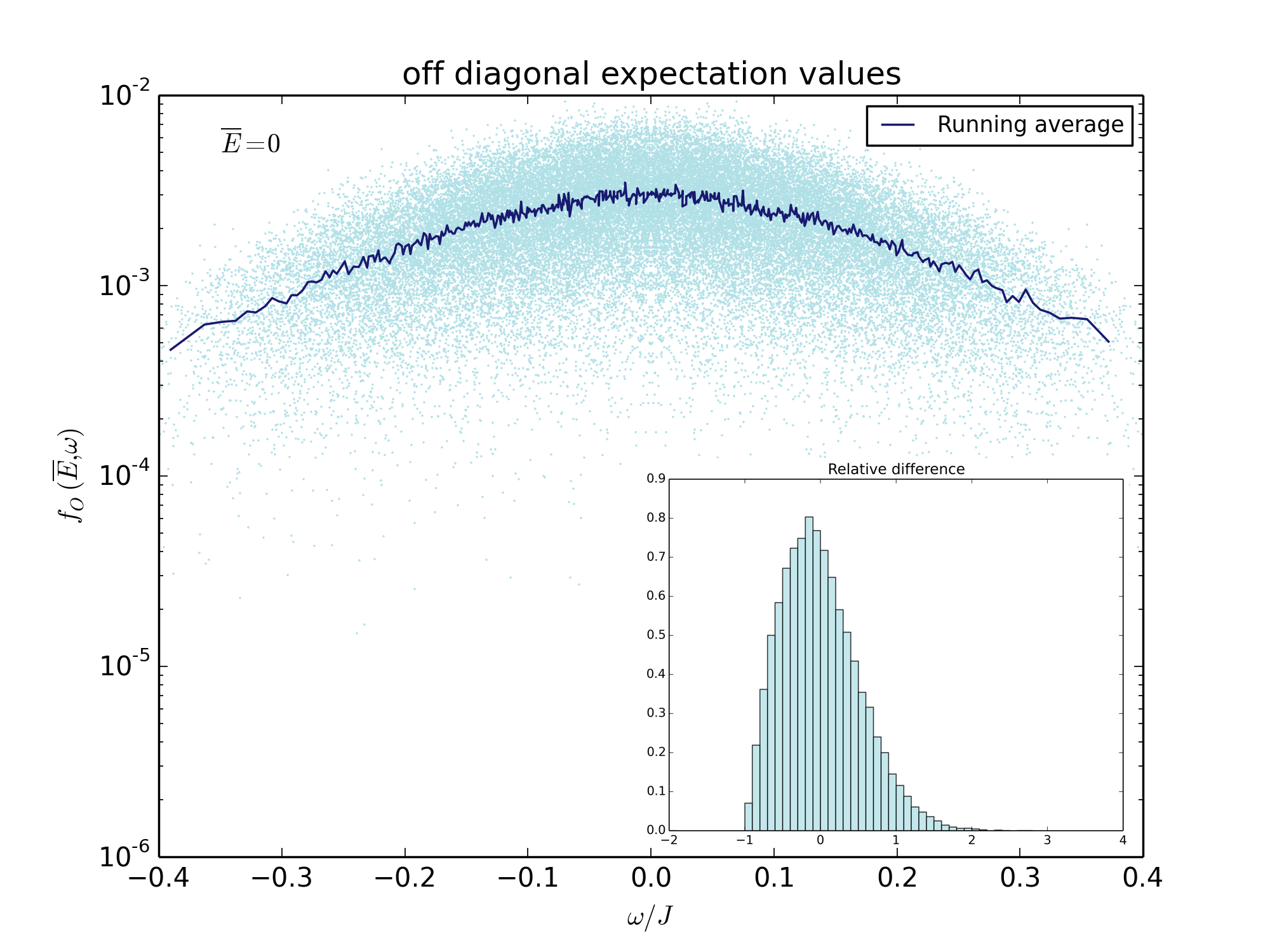} \\ \includegraphics[width=0.48\textwidth]{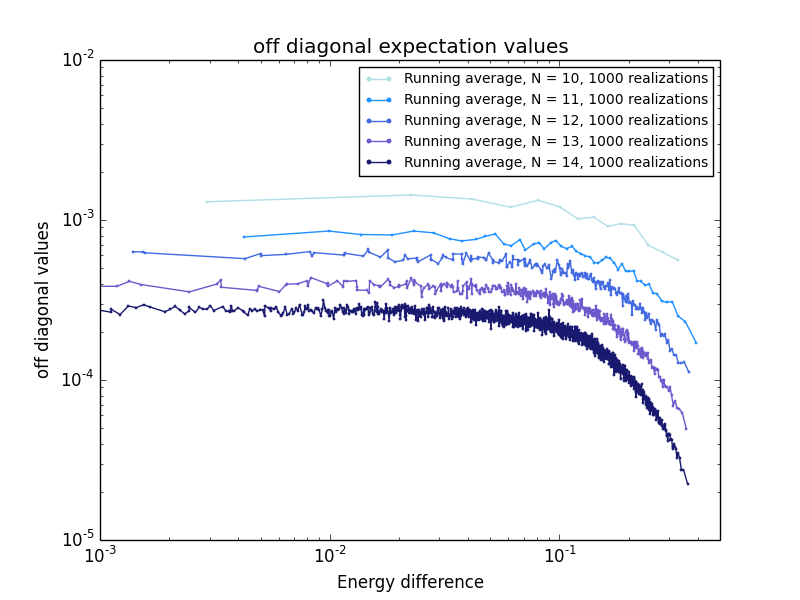} \includegraphics[width=0.48\textwidth]{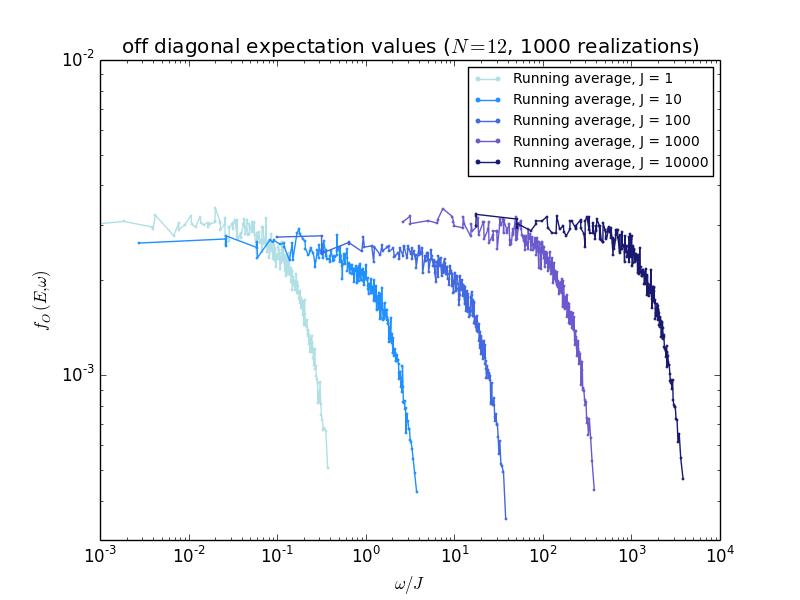}
\caption{\small Off-diagonal values of matrix elements ${\cal O}_{nm}= \left|\langle n | {\cal O} | m \rangle\right| $  for the single-site number operator ${\cal O} = \hat n_N$ at half filling $\nu = \frac{1}{2}$. We show the off-diagonal matrix elements against their energies $E_n/J$. {\bf Top panel}: $N=14$ with raw data in light blue and the running average in dark blue. The inset shows a histogram of relative error between raw data and running average. We see that the histogram is peaked around zero. {\bf Left panel: } the function $f_{\cal O}(\bar E, \omega)$ (we show the running average) for varying Hilbert space dimension corresponding to $N=10,11,12,13,14$. The cross-over from constant to non-constant behavior is identified with the Thouless energy $E_T$. {\bf Right panel: } scaling of the Thouless energy with average coupling strength $J$. A simple fit gives $E_T \propto J^2$.  \label{fig:SYKnfFunction}}
\end{center}
\end{figure}
Moving on to the off-diagonal terms of the matrix elements we demonstrate that the remainder terms are indeed well described by a Gaussian random distribution with zero mean and unit variance, in other words
\be\label{eq.ETHHypothesisOffDiag}
\langle m | \hat{\cal O} | n \rangle = e^{-S(\bar E)/2} f_{\cal O}(\bar E,\omega) R_{mn}\,,\qquad \qquad n\neq m
\ee
In Figure \ref{fig:SYKnMatrixl} we show a histogram of $R_{mn}$ together with a fit to a Gaussian distribution for the single-site number operator. Since the matrix elements are in general complex, both real and imaginary part should be Gaussian distributed and we show a histogram for the imaginary part.
\subsubsection{The function $f_{\cal O}(\bar E,\omega)$}
However, we can go further and calculate the function $f_{\cal O}(\bar E,\omega)$ itself. This captures more detailed physics and allows, for example, to diagnose for what energy ranges the SYK model behaves like a random matrix ensemble, and for what energy ranges it deviates from such behavior. In RMT the function  $f_{\cal O}(\bar E,\omega)$ is a constant function of $\omega$ at fixed $\bar E$ (see Figure \ref{fig:RMTAppendixl} in Appendix \ref{app:RMT}), while deviations become apparent whenever $f_{\cal O}(\bar E,\omega)$ is a non-trivial function of $\omega$. We show the result of this calculation in Figure \ref{fig:SYKnfFunction}, with a cross-over between RMT and non-RMT behavior at a characteristic energy $E_T$, set by the coupling strength $J$. Despite the lack of spatial diffusion in the model we refer to this energy as the Thouless energy $E_T$. Due to this lack of spatial structure (SYK is effectively a zero-dimensional model), the Thouless energy cannot be set by a diffusion time, and it is thus natural that it be set instead by the coupling strength $J$ (at fixed energy $\bar E$). The uppermost panel of Figure \ref{fig:SYKnfFunction} shows a comparison of the raw data with a running average, which is taken over $100$ matrix elements. The resulting smooth curve is then shown in the left bottom panel for a number of different Hilbert space dimensions, as a function of eigenenergy difference $\omega = E_m - E_n$. We can discern a regime over which $f_{\cal O}(\bar E, \omega)$ is almost constant as a function of $\omega$, indicative of RMT like behavior. At a characteristic energy scale $E_T$ this then gives way to non-constant, i.e. non-RMT behavior. In higher dimensional local models this energy is often associated with the diffusive Thouless energy, but, as indicated before, such a physics interpretation appears not to be available in the zero-dimensional SYK case.  We note that this corresponds well with the observations of \cite{Garcia-Garcia:2016mno}, who previously presented evidence for $E_T$ in the Majorana model. We have shown that the scale $E_T$ is set by average strength of the random coupling, $J$, at fixed average energy $\bar E$, in other words it is controlled by the dimensionless coupling $j = J/\bar E$. It should be noted that this is the natural pure-state version of the coupling $\beta J$ which was used in previous studies of the SYK model. We have thus shown that for stronger coupling $j\gg 1$ the range of energies $\omega$ for which the system behaves chaotically (i.e. like RMT) is also increased (see right bottom panel of Figure \ref{fig:SYKnfFunction}). This accords well with intuition as well as previous results in the thermal ensemble indicating chaotic properties to be most pronounced in the strong-coupling regime. Let us now discuss further chaotic aspects of the model.

\section{Correlation functions and chaotic behavior \label{sec.correlationFunctions}}
In section \ref{sec.specTherm} we established the applicability of the ETH ansatz by studying one point functions of nonextensive operators in the complex SYK model. We will now study higher-point correlations in order to elucidate dynamical aspects of the model related to quantum chaos. We will explore how correlation functions in pure states can approximate those in thermal states, relying both on numerics and analytics based on the ETH form on eigenstates. Many of these measures have already been studied in the thermal ensemble\footnote{Where applicable. Of course the spectral form factor, which we study in section \ref{sec:specFormFac} makes no reference to any state or ensemble. At any rate this quantity has already been studied in \cite{Cotler:2016fpe} for the Majorana model and in \cite{Davison:2016ngz} for the spinless Fermion case. }, whereas our focus here is on studying them in pure states. From the holographic dual point of view we are thus investigating the question of how well the correlation functions computed in a black-hole background are approximated by correlations in pure states, in particular in individual eigenstates.

Again, it is interesting to compare the behavior in a single random realization of the model versus the behavior of the same quantity after averaging over a large number of realizations. We shall start, however, with the spectral form factor where the distinction between pure states and the thermal ensemble is meaningless, as can be seen from its definition in terms of an analytically continued partition function. One may also construct the spectral form factor as the fidelity of a certain pure state \cite{delCampo:2017bzr}.

\subsection{The spectral form factor \label{sec:specFormFac}}
The spectral form factor is a well studied quantity in random matrix theory (see e.g. \cite{PhysRevE.55.4067}) as it gives a clean probe of the eigenspectrum of a model. In particular its late-time behavior is sensitive to the discreteness of the spectrum as well as level statistics. The spectral form factor is most conveniently defined in terms of the analytically continued partition function, 
\be
{\cal S}(\beta, t) := \frac{Z(\beta + i t) Z(\beta - it)}{Z(\beta)^2}\,.
\ee
\begin{figure}[t!]
\begin{center}
\includegraphics[width=0.49\textwidth]{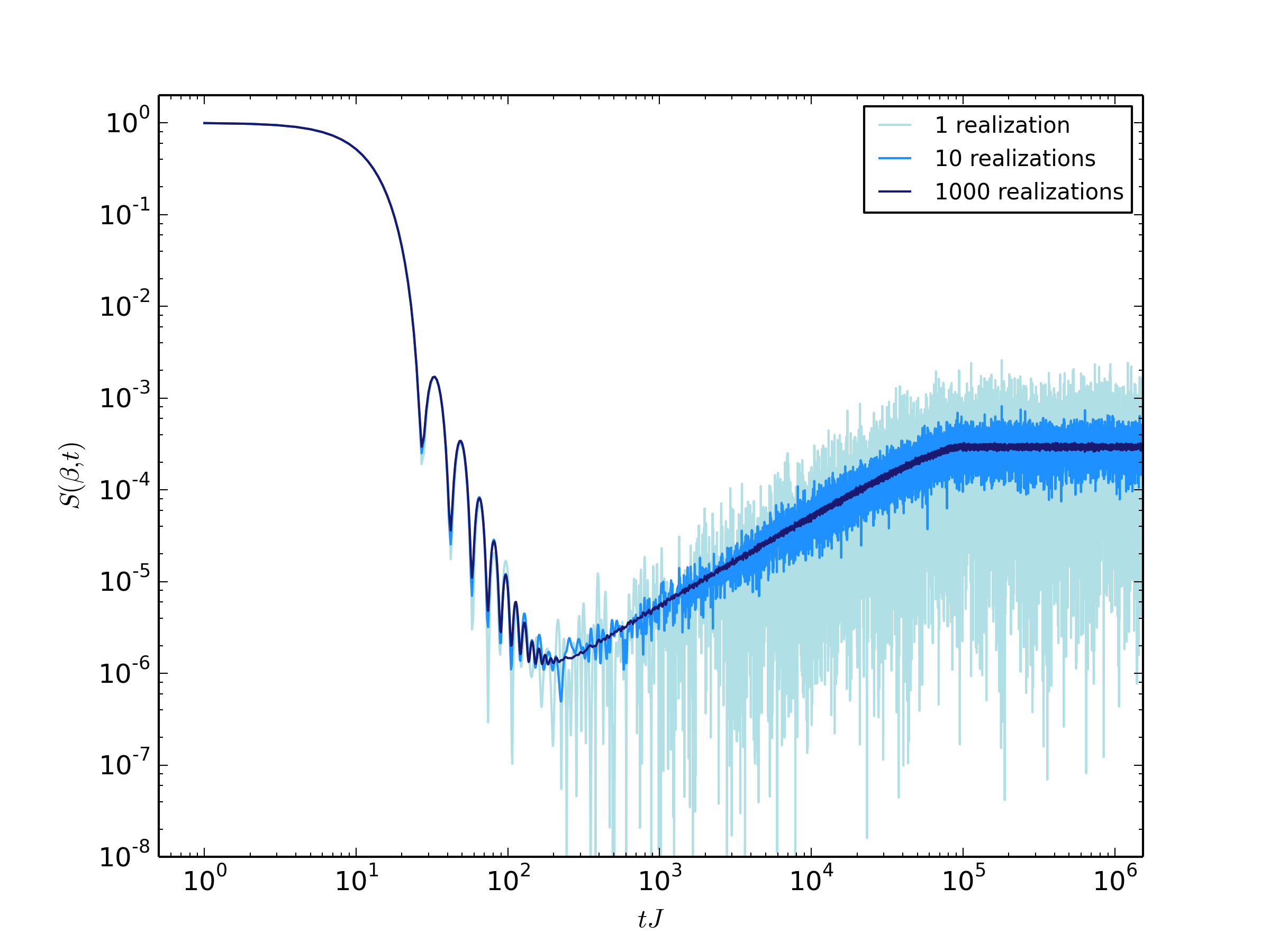} \includegraphics[width=0.49\textwidth]{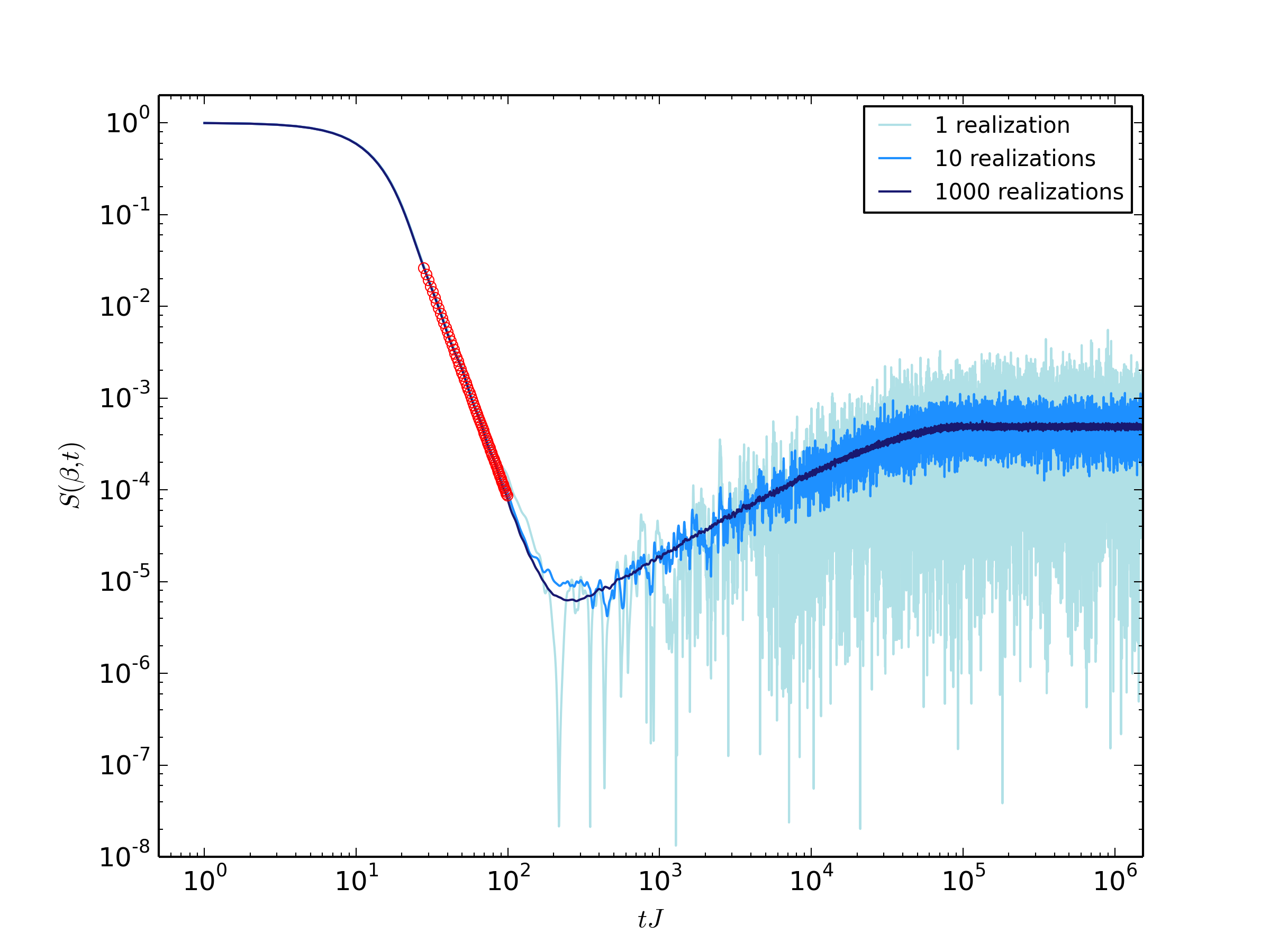}
\caption{\small Spectral form factor at half filling $\nu = \frac{1}{2}$ for varying number of realizations and two different temperatures at $N=14$. In both cases we see the characteristic decay followed by linear ramp and plateau behavior. {\bf Left panel: } $\beta = 1$. One sees several partial revivals in the decaying region with a power-law envelope. {\bf Right panel: } $\beta = 10$. The red markers show points we used for a fit in the slope region, where we find a decay of $\propto t^{-4.53\ldots}$, which is consistent with the value reported in \cite{Davison:2016ngz}. \label{fig:SYKsff}}
\end{center}
\end{figure}
\noindent In Figure \ref{fig:SYKsff} we show ${\cal S}(\beta, t)$ for the spinless Fermion SYK model both as a function of $\beta$ and how it approaches its limiting form as one averages over the random couplings $J_{ij;kl}$. Interestingly we see a $t^{-4.5}$ falloff before the dip time. The fit is shown in red in the right panel of Figure \ref{fig:SYKsff} and the exponent is of course not understood to be exact, depending both on numerical accuracy and the exact choice of the window over which we fit. This does not correspond to the power-law expected from a Wigner edge $\rho(E) \sim \sqrt{E}$ of the spectral density \cite{Cotler:2016fpe}. It is easy to show \cite{delCampo:2017bzr} that a power law ${\cal S}\sim t^{-2(k+1)}$ in the slope region corresponds to a spectral density $\rho(E) \sim E^k$.  At high temperature (left panel) this power law is to be understood as the envelope of an oscillatory decay. Such power laws arise quite generally in computations of survival probabilities of many-body quantum systems \cite{tavora2016inevitable}. We also see the characteristic ramp and plateau behavior  \cite{Cotler:2016fpe, delCampo:2017bzr} at late times which, once more, looks qualitatively like RMT. The difference between RMT and the SYK model is to be found in the precise timescales of dip and plateau times \cite{Cotler:2016fpe}. For completeness and ease of comparison we discuss the RMT spectral form factor in some detail in appendix \ref{app:RMT}.

\subsection{The two-point function}
Let us now turn to the study of correlation functions of non-extensive operators. Here we work with the two-site hopping operator $h_{ij}$ whose one-point functions are studied in Appendix \ref{app:hopping}. The operator is defined as follows: pick two arbitrary sites $i$ and $j$ and write
\be
{\cal O} = h_{ij} = c_i^\dagger c_j + c_j^\dagger c_i\,.
\ee

We have also verified that analogous results hold for the connected and full correlation functions of the on-site number operator $n_i$.
\subsubsection{Eigenstates}
\begin{figure}[h!]
\begin{center}
\includegraphics[width=0.49\textwidth]{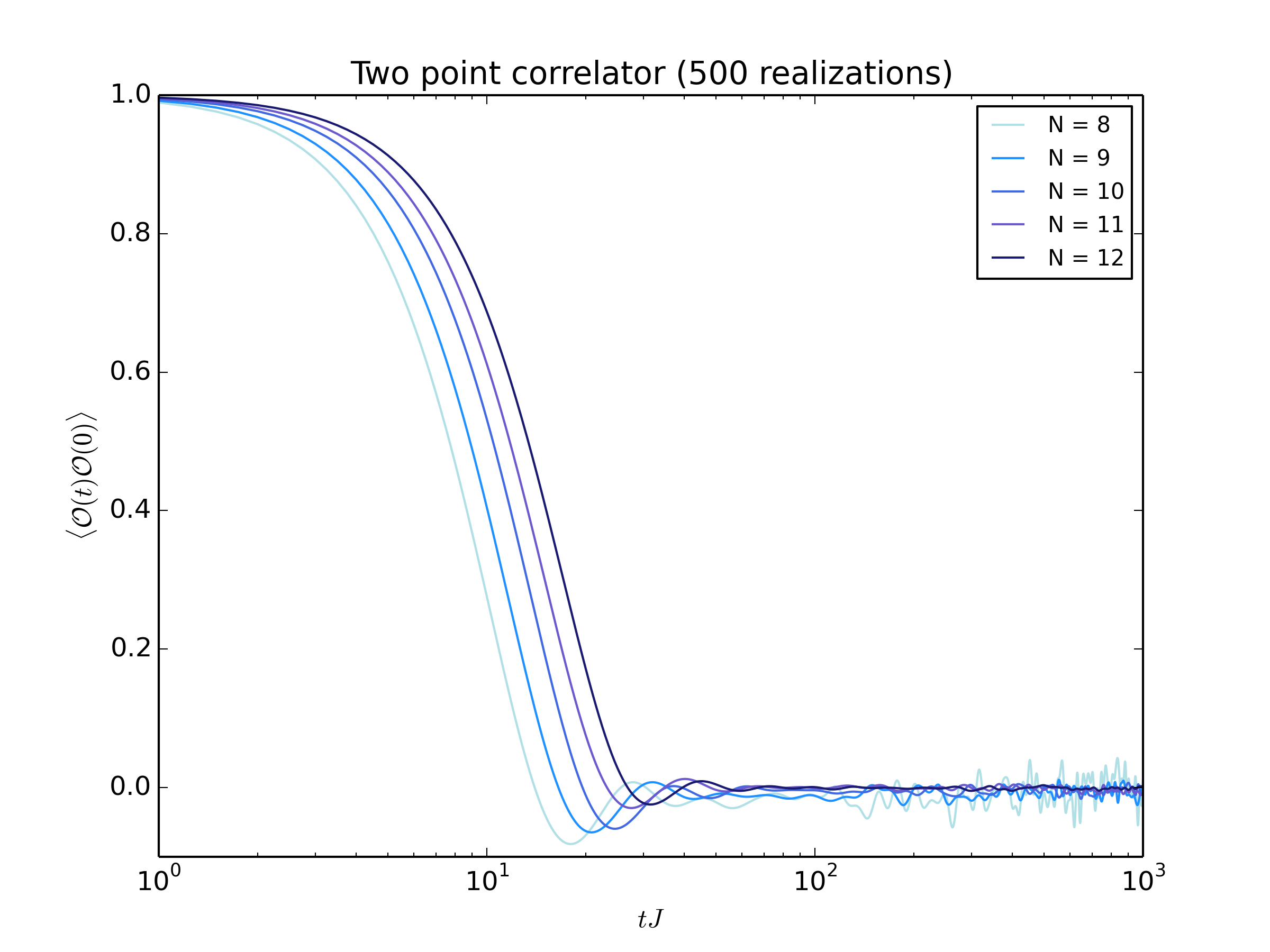} \includegraphics[width=0.49\textwidth]{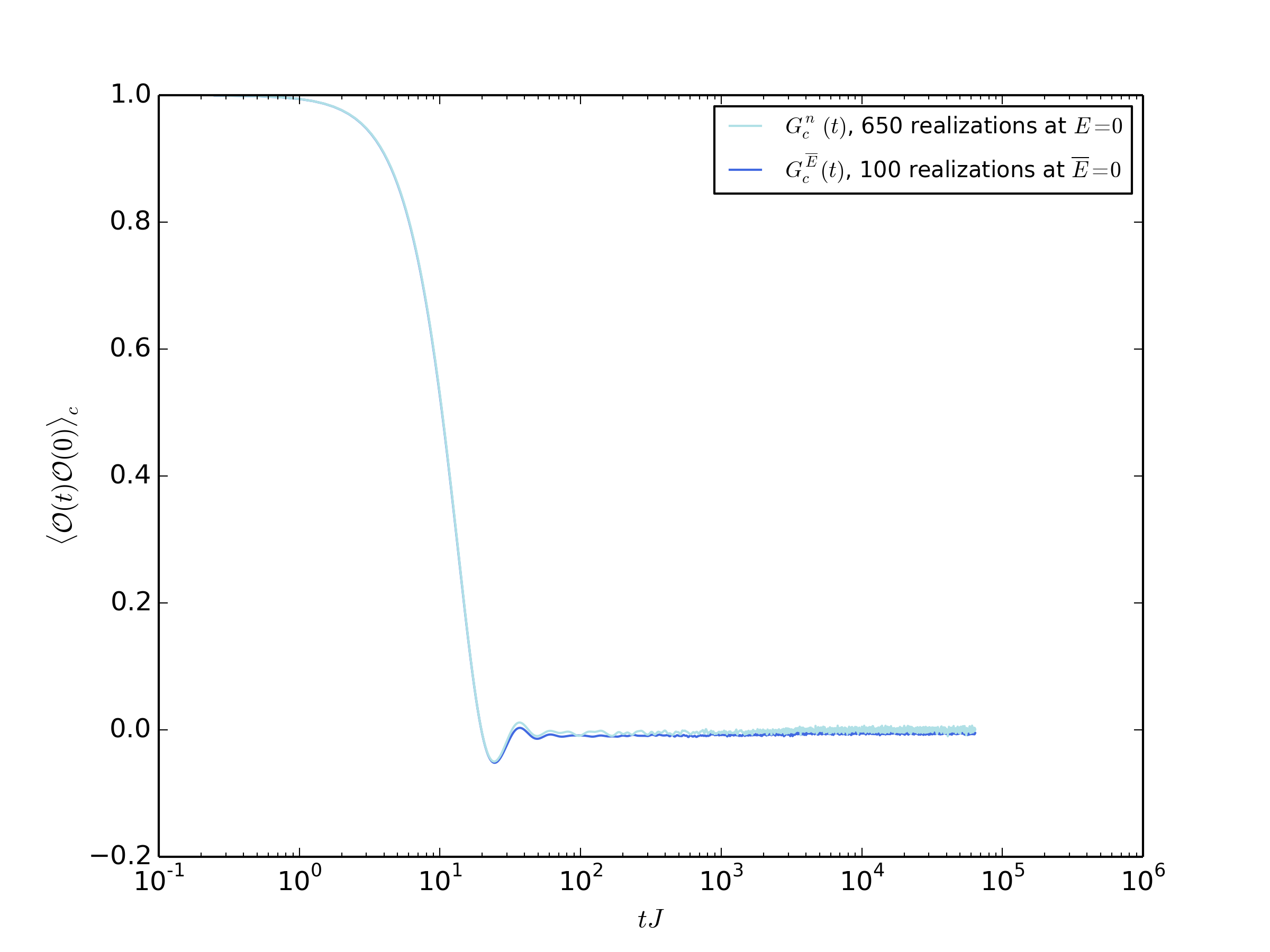}\\
\includegraphics[width=0.49\textwidth]{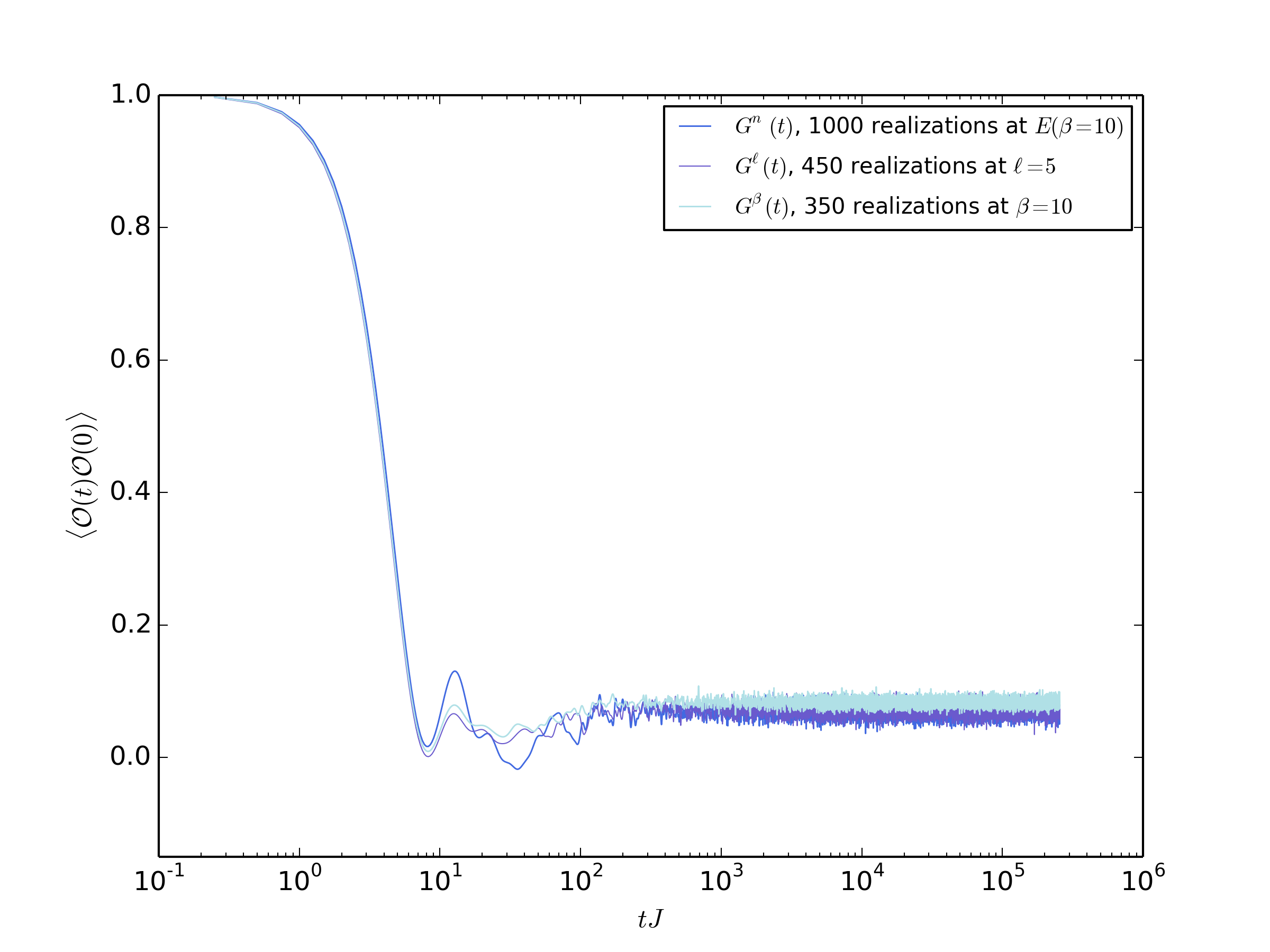} \includegraphics[width=0.49\textwidth]{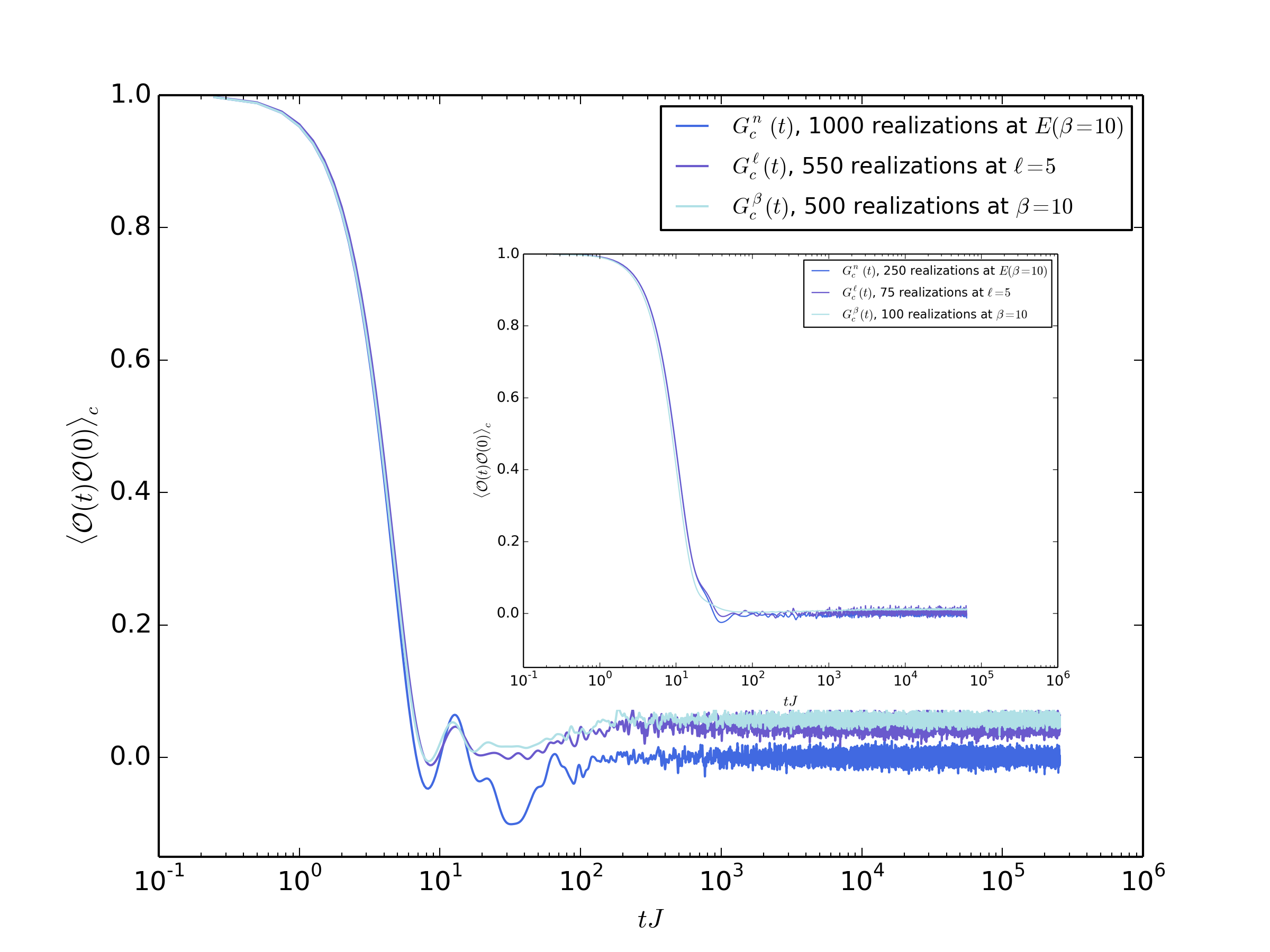}
\caption{\small Two-point function of the hopping operator at half filling $\nu = \frac{1}{2}$.
  {\bf Top left panel: } Two-point correlation function, $G^n_c(t)$, in one eigenstate for $N=8,9,10,11,12$. The initial decay is followed by late time fluctuations around zero of typical size $\sim e^{-S/2}$. {\bf Top right panel: } Comparison of microcanonical $G^{\bar E}_c(t)$ with eigenstate $G^n_c(t)$ at the same energy for $N=10$. One can appreciate the excellent agreement, which would only become better as the number of realizations is increased. {\bf Bottom left panel: } Comparison of $G^\beta(t), G^n(t), G^\ell(t)$  with parameters $\beta,  E(\beta)$ and $2\ell = \beta$ at $N=6$. {\bf Bottom right panel: }  Comparison of $G^\beta_c(t), G^n_c(t), G^\ell_c(t)$  with parameters $\beta$, $E(\beta)$ and $2\ell = \beta$ at $N=6$ (inset $N=10$).  \label{fig:SYK2ptI}}
\end{center}
\end{figure}

We now study correlation functions of the hopping operator in energy eigenstates. For definiteness we will take sites $i=N-1$ and $j=N$, but any two sites will give essentially the same answer. We consider the two point function of the hopping operator,
\be
G^{n}(t) =  \langle h_{ij}(t) h_{ij}\rangle_{E_n} := \langle n|e^{iHt}h_{ij}e^{-iHt}h_{ij} |n\rangle
\ee
for some excited eigenstate $|n\rangle$ with energy $E_n$, as well as its connected cousin
\be
G^n_c(t) =  \langle h_{ij}(t) h_{ij}\rangle_{E_n}  - \langle h_{ij}\rangle  \langle h_{ij}\rangle_{E_n}\,.
\ee 
We find that the connected correlation function in eigenstates quickly decays and subsequently oscillates around zero as shown in Figure \ref{fig:SYK2ptI}. This latter fact is easily established by averaging over time. One finds
\be\label{eq.eigenstateAverageLateTime}
\overline{G^n(t)} = \lim_{T\rightarrow \infty}\frac{1}{T} \int_0^T  \sum_me^{i(E_n - E_m)t}|{\cal O}_{nm}|^2dt = |{\cal O}_{nn}|^2\,,
\ee
where we temporarily denoted the Hermitian operator $h_{ij}$ by ${\cal O}$ to avoid too much index clutter. It follows that the connected eigenstate correlation function averages to zero, $\overline{G^n_c(t)}  = 0$, at late times. The typical size of the late-time fluctuations follows from eigenstate thermalization, (\ref{eq.ETHHypothesis}), to be $\sim e^{-S/2}$. We show explicit computations of $G^n_c(t)$ for different values of $N$ in Figure \ref{fig:SYK2ptI} (top left).

As implied by (\ref{eq.ETHHypothesis}), the behavior of the two-point function in eigenstates approximates very closely the corresponding microcanonical quantities. This agreement is expected to be perfect in the thermodynamic limit $N\rightarrow \infty$. By a microcanonical two point function of an operator $\cO$ we mean the quantity
\be
G^{\bar E}(t) = \frac{1}{{\cal N}_{\bar E}} \sum_n \langle  n | \cO(t) \cO |n \rangle\,,\qquad\qquad \overline{G^{\bar E}(t)} = \frac{1}{{\cal N}_{\bar E}}\sum_n |\cO_{nn}|^2,
\ee
where ${\cal N}_E$ is the number of states in a window of energies of given width $\Delta E$ around the average energy $\bar E$ and the sum over $n$ runs over exactly those states. This agreement is illustrated in Figure \ref{fig:SYK2ptI} (top right). As a basic check one can convince oneself that (\ref{eq.ETHHypothesis}) implies that its long time average gives exactly (\ref{eq.eigenstateAverageLateTime}), which is also borne out in Figure  \ref{fig:SYK2ptI} (top right). We conclude therefore that two-point functions in the disorder-averaged theory become arbitrarily close to their thermal (microcanonical) averages. This agreement is perhaps not surprising if one realizes that averaging the correlation function over couplings is operationally similar to a microcanonical average in the first place. 

However, we now want to compare the behavior of $G^n$ and $G^n_c$, to the corresponding thermal correlation functions with respect to the canonical density matrix $\rho = e^{-\beta H}$, at energy $E(\beta)$, determined by the map (\ref{eq.canonicalEnergyTemperatureMap}), that is
\be
G^\beta (t) =\frac{1}{Z(\beta)} {\rm Tr}\left[ e^{-\beta H}h_{ij}(t) h_{ij} \right]
\ee
and its connected cousin
\be
G^\beta_c (t) = G^\beta (t) - \frac{1}{Z(\beta)^2} \left({\rm Tr}\left[ e^{-\beta H} h_{ij} \right]\right)^2,
\ee
where again $h_{ij}(t)$ is the hopping operator in the Heisenberg picture at time $t$. Generally speaking all these correlation functions show the expected behavior, namely an early time exponential decay, followed by intermediate-time power law decay, further followed by a late time ramp to a very late time plateau. The plateau value may be zero, or non-zero depending on which operator and which correlation function one considers.

As illustrated in Figure \ref{fig:SYK2ptI} (bottom left) we find that the full correlation function $G^n(t)$ in eigenstates at energy $E_n(\beta)$ starts to approximate the corresponding thermal one $G^{\beta}(t)$ at early time and at late times. Here $E(\beta)$ is the energy corresponding to the inverse temperature $\beta$ via the map (\ref{eq.canonicalEnergyTemperatureMap}), while at intermediate times (during the ramp), $G^n(t)$ and  $G^{\beta}(t)$ can differ.

The connected correlation function $G^n_c(t)$ in eigenstates oscillates around zero at late times, unlike its thermal equivalent $G^{\beta}_c(t)$, which oscillates around a non-zero average value as seen in Figure \ref{fig:SYK2ptI} (bottom right). These differences are subleading in the size of the Hilbert space and are expected to become negligible in the $N\rightarrow \infty$ limit. The latter limit is of special interest for a putative gravitational dual as it corresponds to the semi-classical regime where a geometric description should become possible. 

As a side comment, in the limit $\beta\rightarrow\infty$, the thermal expectation value starts approximating the eigenstate one arbitrarily closely. This, of course, has nothing to do with thermalization, as it corresponds to the zero-temperature limit, where the `thermal' average projects on the ground state, and is thus manifestly equal to the eigenstate correlation function. 
In conclusion then we find that the two-point function in individual eigenstates of the disorder-averaged theory behaves thermally, showing the most precise match with the microcanonical average of the correlation function. Since we work at finite $N$ the different statistical ensembles need not give the same answers, and indeed subtle differences are seen between canonical and eigenstate correlation functions. These differences are expected to disappear in the thermodynamic limit. We can already appreciate the convergence of the different ensembles for $N=6$ versus $N=10$ by comparing Fig. \ref{fig:SYK2ptI} (bottom right) with its inset.

\subsubsection{Superposition states}
Up to this point we have considered mostly eigenstates. From what we have found one can conclude also that arbitrary pure states with narrow support in energy thermalize to microcanonical averages at late times, consistent with eigenstate thermalization. However, thermalization of states with broad support in energy do not thermalize in this way. We will next consider pure states, closely related to the ones considered in \cite{Kourkoulou:2017zaj}, with very broad spread over the energy spectrum\footnote{For a discussion of similar states in the context of tensor models, see \cite{Krishnan:2017txw}} and demonstrate that they nevertheless thermalize, but more precisely to canonical averages. Note, again, that at finite $N$  microcanonical and canonical averages do not have to exactly agree, and consequently one or the other may be a better approximation to a thermalizing pure state correlation function. 

We consider pure states which are superpositions of eigenstates, as one would obtain, e.g. as a result of a sudden quench. An interesting class of such pure states can be constructed as follows. Let $|{\cal C}\rangle$ be a canonical state at half filling. Select a set of $N_{\cal C}$ creation operators $S_{\cal C} =\left\{c_a^\dagger \right\}_{a=1}^{N_{\cal C}}$. Then let
\be
|{\cal C}\rangle = \prod_{i\in S_{\cal C}}c_i^\dagger | \Omega \rangle\,,
\ee
where $|\Omega\rangle$ is the state with no fermions. We assumed that $N$ is even and we work at half filling $\nu = \tfrac{N_{\cal C}}{N} = \tfrac{1}{2}$, but such states can be constructed also for odd $N$ and other filling fractions. What is important is that we think of these states at large $N$ as being finitely excited, i.e. $\nu$ is held fixed as $N\rightarrow \infty$.
Now we define a one-parameter family of pure states via Euclidean evolution of the canonical state,
\be
|\ell \rangle = e^{-\ell H}|{\cal C}\rangle\,.
\ee
Such a state can be expanded in the eigenbasis
\be
|\ell \rangle  = \sum_\alpha e^{-\ell E_\alpha} c_\alpha |E_\alpha\rangle
\ee
where, according to ETH, the coefficients $c_\alpha$ are Gaussian distributed complex random variables \cite{srednicki1994chaos}.  We expect these states to behave approximately thermally at a temperature $\beta^{-1}$ that we can determine by the map (\ref{eq.canonicalEnergyTemperatureMap}), together with the fact that the eigenstates satisfy (\ref{eq.ETHHypothesis}). To this end we compute
\be
\langle \ell |H|\ell \rangle = \sum_{\alpha,\beta} c_\alpha^* c_\beta e^{-\ell(E_\alpha + E_\beta)}E_\alpha\,.
\ee
The random nature of the coefficients $c_\alpha$ ensures that this expectation value behaves like a thermal average. We can make this more precise by averaging the expansion coefficients over the eigenstate ensemble \cite{srednicki1994chaos}, using
\be
\left[c_\alpha^* c_\beta \right] = \frac{1}{N_{\cal C}}\delta_{\alpha\beta}\,.
\ee
With the help of this expression, we find
\be
\frac{\langle \ell |H|\ell \rangle}{\langle \ell |\ell \rangle}\longrightarrow \frac{1}{Z(2\ell)} \sum_{\alpha}e^{-2\ell E_\alpha}E_\alpha\,,
\ee
where we have averaged numerator and denominator independently.
This quantity is equivalent to the thermal expectation value $\langle  E\rangle_{\beta = 2\ell}$ since the state $|{\cal C}\rangle$ typically has support over the whole spectrum. It is thus natural to compare expectation values in $|{\cal C} \rangle$ with thermal expectation values in the canonical ensemble at $\beta = 2\ell$. The behavior of these states in a sense is similar to the thermofield double state, with the role of the trace over the second copy taken over by the random distribution of expansion coefficients. In Fig (\ref{fig:SYK2ptI}) we show the correlation function
\be
G^\ell(t) = \langle \ell | h_{ij}(t) h_{ij}|\ell \rangle
\ee
and its connected version $G^\ell_c(t)$, defined in the obvious way, in comparison with the analogous canonical averages. We see that in fact they are rather close to their thermal counterparts at inverse temperature $\beta = 2\ell$ as expected. Let us next move on to four-point functions, an in particular the issue of chaos in eigenstates.
\subsection{The four-point function}\label{sec.OTOC}
\begin{figure}[h!]
\begin{center}
\includegraphics[width=0.52\textwidth]{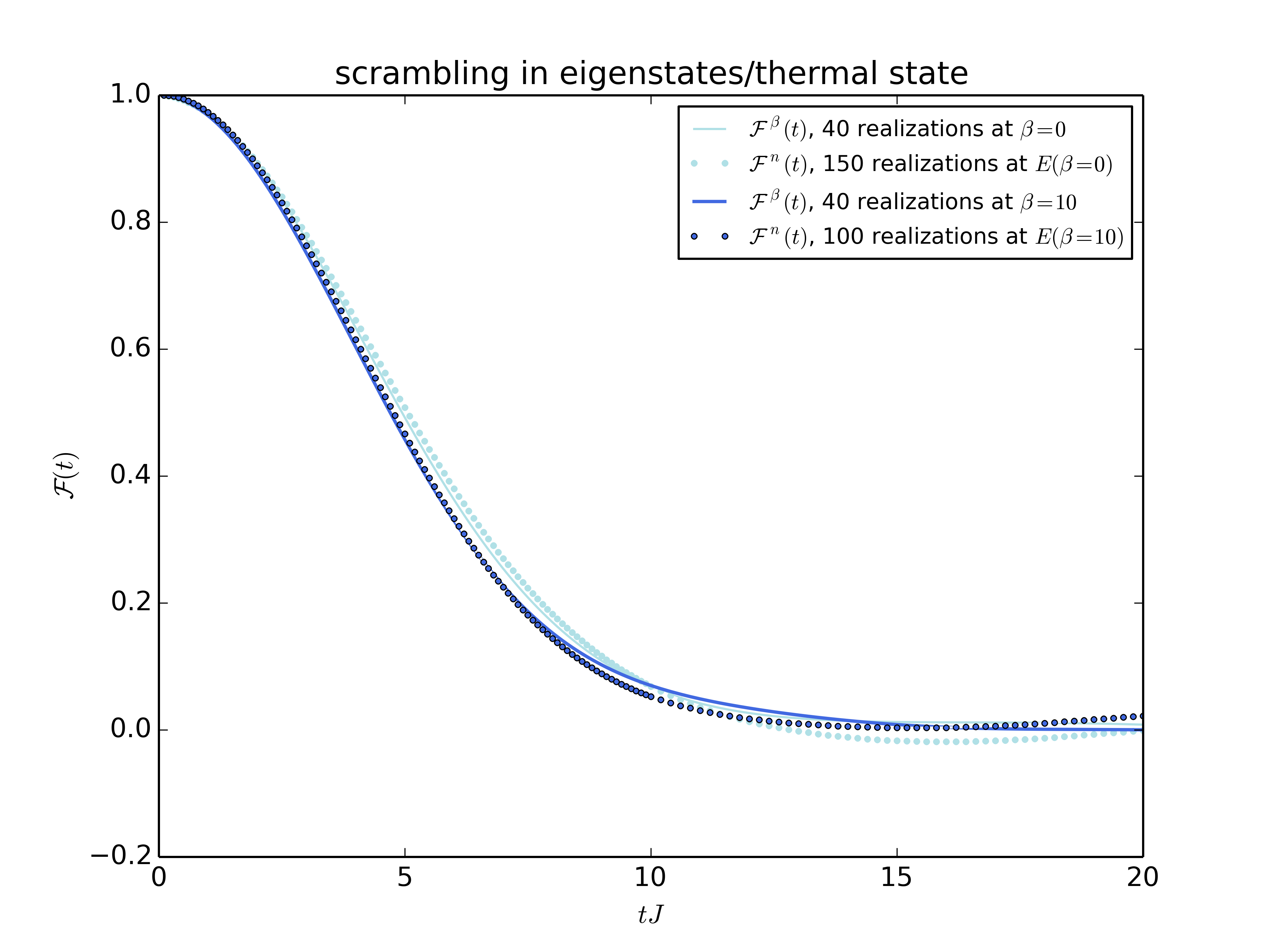}\includegraphics[width=0.52\textwidth]{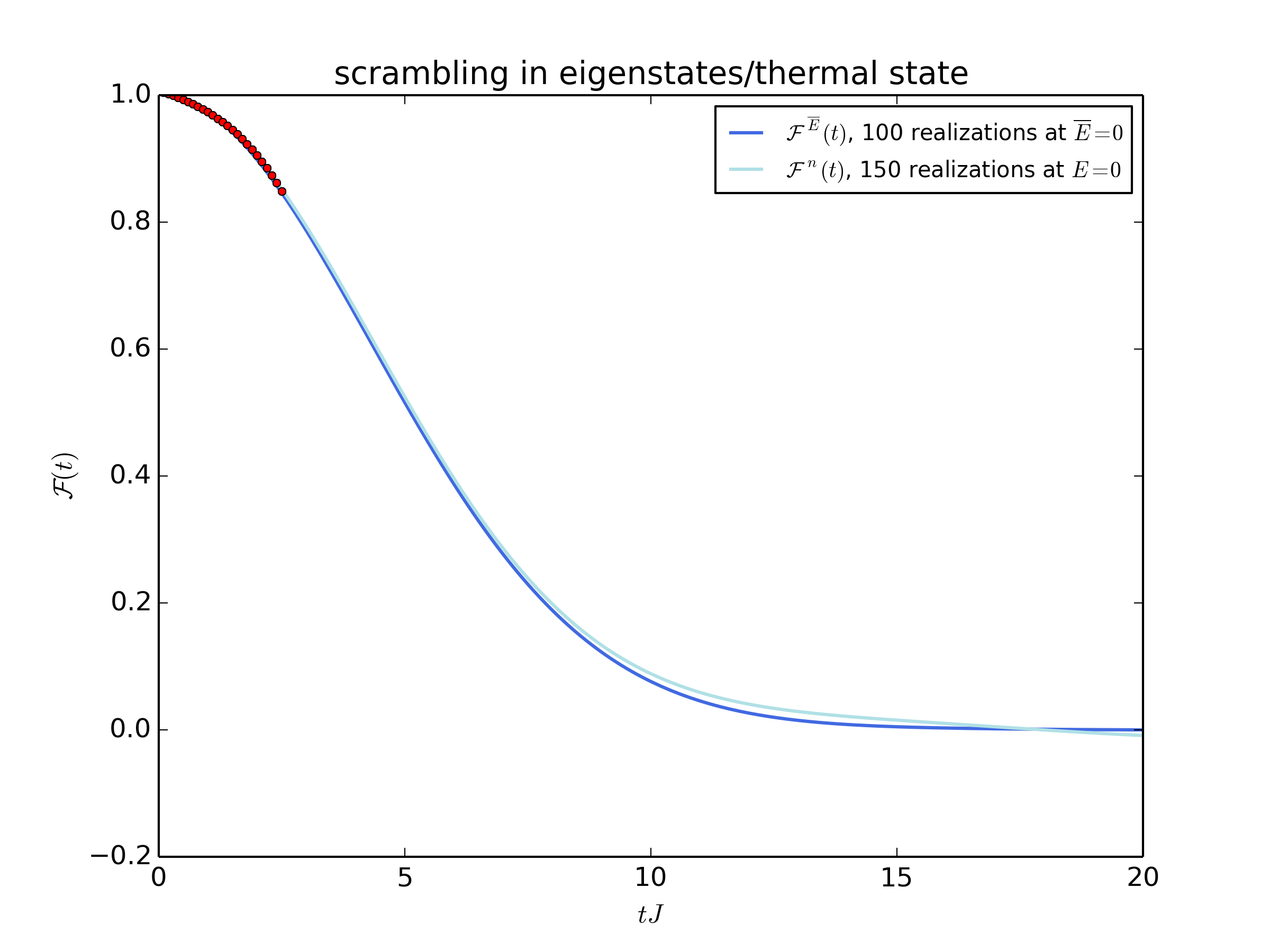}
\caption{\small Four-point out-of-time order (OTO) correlation function of the hopping operator.  {\bf Left panel: } Four-point OTO correlation function in individual eigenstates vs. thermal OTO correlation function at temperature $\beta(E)=0,10$ (for $N=7$). {\bf Right panel: } Four-point OTO correlation function in individual eigenstate $|n\rangle$ vs. thermal OTO correlation function in the microcanonical ensemble at energy $\bar E = E_n$ (for $N=7$) with a fit to the functional form (\ref{eq.EstateLyapunov},\ref{eq.OTOCScramble}) for $\alpha = 0.033\ldots$ and $\lambda = 0.69\ldots$ .\label{fig:SYK4ptI}}
\end{center}
\end{figure}
We now finally turn to studying the four-point function\footnote{The results in this section were obtained in collaboration with Jérémie Francfort.} which serves to characterize early time chaos via the Lyapunov exponent $\lambda_L$ defined in equation (\ref{eq.estateLyapunov}). Again we emphasize that this has been studied extensively in the thermal ensemble \cite{Maldacena:2015waa,AK15,Fu:2016yrv,Maldacena:2016hyu}, while here our main focus is to study it in eigenstates.

We again focus on the two-site hopping operator $h_{ij}$. For an SYK model on a grid of size $N$, let us define two operators
\be
W = h_{N-1,N}\qquad V = h_{N-3,N-2}
\ee
 together with their out-of-time-order four-point function
\be
{\cal F}(t) = \frac{\langle W(t) V(0) W(t) V(0)\rangle+  \langle V(t) W(0) V(t) W(0)\rangle}{2\left\langle W(0)W(0)V(0)V(0)  \right\rangle}\,.
\ee
The expectation value is taken, as before, in a finite-energy eigenstate, denoted ${\cal F}^n(t)$, or for comparison in the (micro-)canonical ensemble, denoted ${\cal F}^{\bar E}(t)$, ${\cal F}^\beta(t)$, respectively. The expected form (see (\ref{eq.EstateLyapunov})) of this function at times up to the scrambling time is
\be\label{eq.OTOCScramble}
{\cal F}(t) = {\cal F}_0 - \alpha e^{\lambda t},
\ee
where the coefficient $\alpha = \beta J/N$ in the canonical ensemble, at strong coupling and large $N$ \cite{Maldacena:2015waa}, the same circumstances under which the Lyapunov exponent takes its maximal value $\lambda = 2\pi/\beta$ \cite{AK15,Maldacena:2015waa}. For our OTO correlation functions ${\cal F}^{\bar E}(t)$, which well approximates that in the corresponding eigenstate, we show a fit to the form (\ref{eq.OTOCScramble}) in Figure \ref{fig:SYK4ptI}. Our results here are consistent with those of \cite{Fu:2016yrv} who considered ${\cal F}^\beta(t)$ in exact diagonalization and concluded that at finite $N$, the Lyapunov exponent is not maximal and does not vary inversely with $\beta$, but rather that the parameter governing $\lambda$ is the coupling $J$. The behavior of the OTO correlator in eigenstates thus accords very well with expectations from eigenstate thermalization. In particular the early-time physics, including the scrambling time, of this correlator in eigenstates coincide to numerical accuracy with the thermal results. This suggests that the large-$N$ OTO correlator in eigenstates ${\cal F}^n(t)$ will also take the form (\ref{eq.OTOCScramble}) with Lyapunov exponent given by (\ref{eq.EstateLyapunov}).

\section{Discussion}
In this work we have endeavoured to establish the mechanism of thermalization in the complex spinless SYK model, as a toy model of a strongly correlated quantum system with a holographic dual\footnote{\label{foot:bulkDual}To the extent that such a dual has been established. See \cite{Jackiw:1984je,Teitelboim:1983ux,Almheiri:2014cka,Jensen:2016pah,Maldacena:2016hyu,Maldacena:2016upp,Mandal:2017thl,Taylor:2017dly,Das:2017pif} for extensive discussions and results on this issue, including an explicit brane construction whose spectrum contains the exact SYK spectrum \cite{Das:2017pif}.}. We focused on the complex model, but we expect the Majorana model to exhibit qualitatively similar behavior, that is to say that we believe it also satisfies eigenstate thermalization.

 We were careful to establish eigenstate thermalization both for an individual random realization - with increasing accuracy as the Hilbert space dimension $2^{N}$ is increased -- as well as for a fixed Hilbert space dimension at moderate $N$ -- with increasing accuracy as one averages over more and more realizations. 

We have also studied the extent to which two-point and four point correlations in finite-energy pure states approximate those in the thermal ensemble at the corresponding temperature. Our results support the conclusion that individual eigenstates in the SYK model behave thermally. We established that the agreement between pure and thermal expectation values becomes better for a single realization of the model as $N$ is increased and for a fixed finite $N$, as we average over a larger and larger ensemble of Hamiltonians. However, consistent with earlier findings \cite{Cotler:2016fpe} we observe that correlation functions are self-averaging at early times, but lose this property at late times. This property is shared by the spectral form factor. Herein lies an important subtlety: a bulk dual\footnote{See previous footnote.} of SYK has been proposed for the disorder averaged theory, which means that any bulk solution is strictly dual to an ensemble of boundary Hamiltonians. One should therefore not associate a single eigenstate of an individual realization of the boundary theory with the late-time behavior of a bulk solution. This point does not apply to the tensor models of \cite{Witten:2016iux,Klebanov:2016xxf,Gurau:2016lzk}. It should, however, be kept in mind during the rest of the discussion section.
\subsection{Comments on putative bulk dual}
Let us now address the question of the interpretation of our results in terms of a putative holographic dual, keeping in mind our previous comments on the status of such a dual. A crucial issue in the holographic description of black holes is the representation of their interior from the boundary theory point of view \cite{Papadodimas:2013wnh}. One application of our results in this respect concerns the relationship between entanglement and geometry \cite{Maldacena:2001kr,Maldacena:2013xja} (see also \cite{Jensen:2013ora,Sonner:2013mba} for a world-sheet analog). One may appeal to the approach of \cite{Marolf:2013dba} to argue that a typical highly entangled eigenstate of the SYK model does not have a dual with a smooth geometrical connection. The argument of \cite{Marolf:2013dba} uses eigenstate thermalization as a hypothesis to roughly reason as follows. We note that a typical two-sided correlation function in the eternal black hole geometry will be of order $e^S$ at early times, coming from the wormhole connecting the two boundaries \cite{Maldacena:2001kr,Maldacena:2013xja}. One then appeals to the eigenstate expectation values  of the form (\ref{eq.ETHHypothesis}) to argue that the same operator in a generic highly entangled finite energy state does not have the required $e^S$ correlations at early time, in fact it is exponentially suppressed. This way one arrives at a contradiction with the assumption that the correlator can be computed in a smooth geometry with a wormhole connection between the two boundaries. Closely related ideas have been advanced in \cite{Balasubramanian:2014gla}. By establishing eigenstate thermalization in the SYK/NAdS$_2$ context, one important implication of our work is that a generic highly entangled state of the SYK model either does not have a smooth geometric dual, or that entanglement does not generically correspond to having a geometrical wormhole in the putative bulk dual of SYK. However,  without entering into the details, if one allows the state-dependent construction of interior operators by \cite{Papadodimas:2012aq}, smooth black hole interiors may be constructed. 

Of course more directly eigenstate thermalization tells us that one-sided correlation functions look thermal even in eigenstates. This means that correlations in individual eigenstates are well approximated by dual computations in the black hole geometry. Similar results apply in AdS$_3$/CFT$_2$ see e.g. \cite{Fitzpatrick:2015foa,Fitzpatrick:2015zha}, where two-point functions of light operators in states created by heavy primary operators were shown to be well approximated by the corresponding results in the BTZ black hole and non-equilibrium initial states thermalize exactly to this state \cite{Anous:2017tza,Anous:2016kss}.

 As already alluded,  eigenstate thermalization has been discussed also as the mechanism of thermalization in two-dimensional CFT \cite{Lashkari:2016vgj,Basu:2017kzo,Fitzpatrick:2015zha,Fitzpatrick:2015foa} as well has higher-dimensional cases \cite{Lashkari:2016vgj}. In higher dimensions a direct approach, such as in this paper, seems out of reach. It would therefore be interesting to gain a more analytical understanding of our results, and we hope to address this in the future. It will be interesting to try and carry our results over into a more widely applicable picture of thermalization in theories with holographic duals. In this respect it may be interesting to map out the applicability of eigenstate thermalization in more SYK-like models, such as \cite{Anninos:2016szt,Fu:2016vas,Li:2017hdt,Berkooz:2016cvq}. Conversely, if one instead accepts that eigenstate thermalization should hold in theories with holographic duals, one might hope to use the constraints on matrix elements due to eigenstate thermalization, in order to further refine the requirements on CFTs with a well-behaved holographic dual. 

 It would be interesting to repeat the study in the present paper using the tensor models of \cite{Witten:2016iux,Klebanov:2016xxf,Peng:2016mxj}  and in particular to study how much of what we uncover here survives away from the large-$N$ semi-classical limit in such theories, which have the advantage of being defined without the quenched disorder average. Certain spectral and chaotic properties have been studied via exact diagonalization  by \cite{Krishnan:2016bvg}. 
 
In conclusion we believe that the detailed study of thermalization via eigenstates in SYK, both numerically and analytically, gives us a concrete opportunity to better understand the physics of quantum black holes, at least at the level of toy models.

\section*{Acknowledgements}
We would like to thank Bela Bauer for his early invaluable help in writing an efficient numerical algorithm for the complex SYK model. We thank  Dmitry Abanin, Shouvik Datta, Thierry Giamarchi, Wen Wei Ho, Pranjal Nayak, Kyriakos Papadodimas, Eliezer Rabinovici, and Benjamin Withers for discussions. Our special thanks go to Jérémie Francfort for his collaboration at an early stage. The simulations in this paper were performed on the baobab HPC cluster at the University of Geneva and we thank Yann Sagon for his assistance.

\appendix
\section{Full spectrum \& ETH for other operators}\label{app:hopping}
This appendix is concerned with filling in some details on the spectrum of the model, as well as to supply more details on eigenstates thermlization for a different non-extensive operator, namely the hopping operator.
\begin{figure}[h!]
\begin{center}
\includegraphics[width=0.49\textwidth]{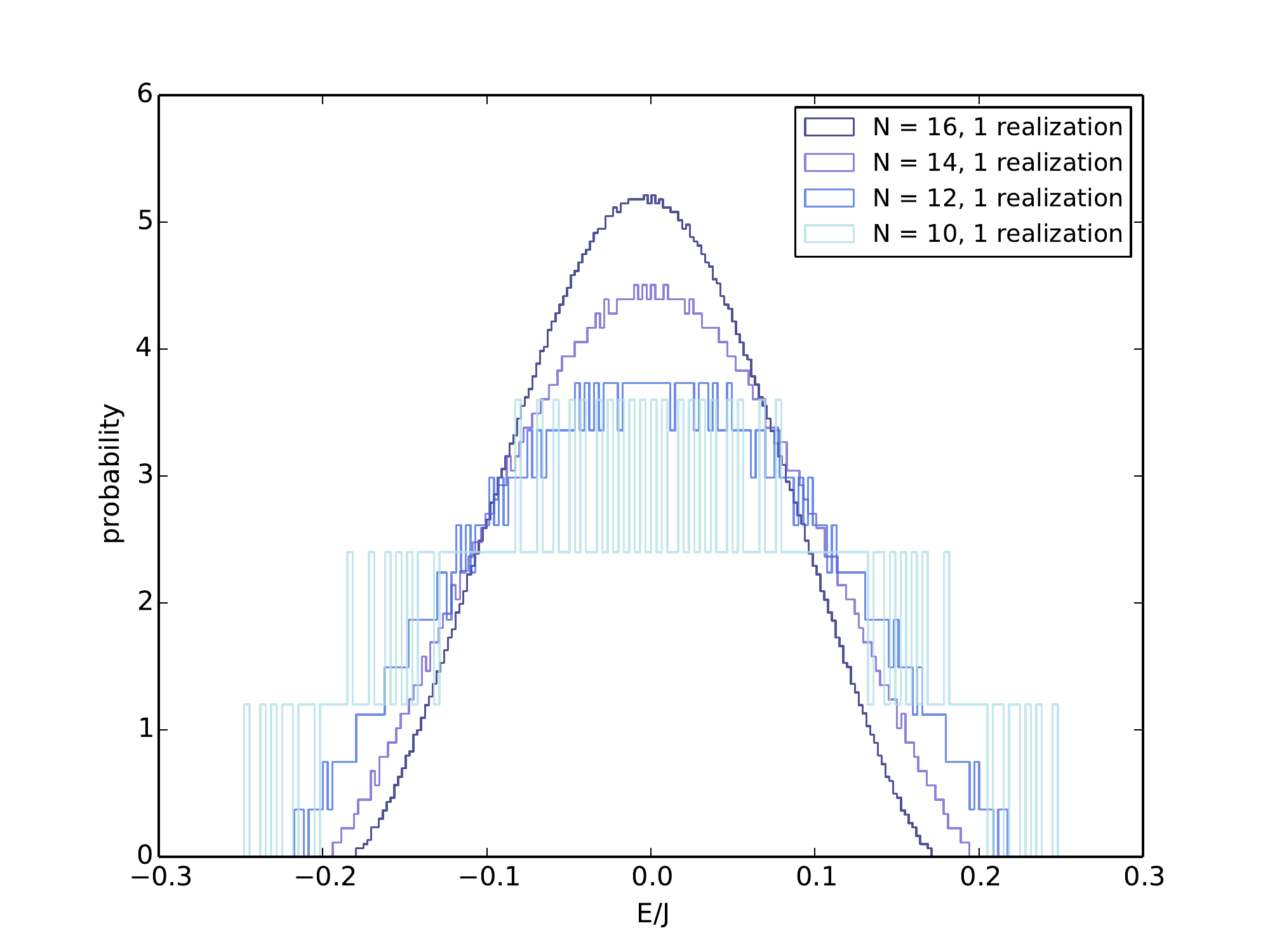} \includegraphics[width=0.49\textwidth]{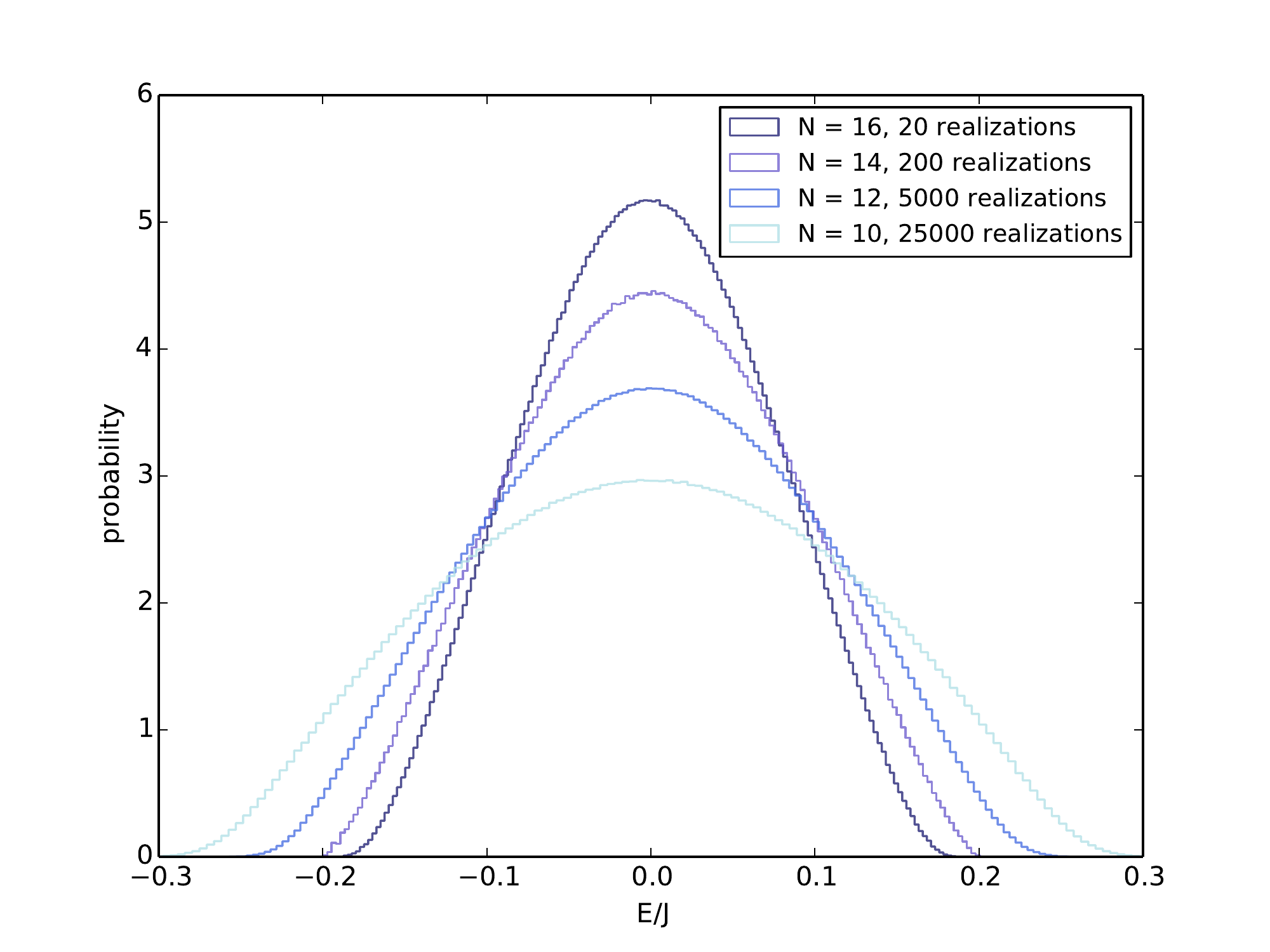} 
\caption{\small Unit normalized histogram of energy eigenvalues (in units of $J$) in the SYK model for $N=10,12,14,16$ sites. This converges to the continuum probability density $\rho(E)$. {\bf Left panel}: spectrum for a single realization. {\bf Right panel}: spectrum averaged over disorder realizations. We see that already for $N=16$ the spectrum for a single realization is essentially indistinguishable from the ensemble average.  \label{fig:SYKspectrum}}
\end{center}
\end{figure}
A careful analysis of the spectral properties of the SYK model was carried out in \cite{Polchinski:2016xgd,Maldacena:2016hyu,Garcia-Garcia:2016mno,Cotler:2016fpe,Garcia-Garcia:2017pzl}. Here we present some essential features on the complex-spinless case, to set the context, and also to benchmark our numerics. More details are presented in the aforementioned references.
\subsection{Density of states}

The full spectrum of the model is most efficiently computed by considering each allowed filling fraction $\nu$ separately. It is both of interest to consider the spectrum of a single randomly chosen realization as well as averaged over a large number of realizations (see. Figure \ref{fig:SYKspectrum}). As has been observed before, for the Majorana model, the spectrum self averages very well even for moderate values of $N$ as can be surmised from comparing left and right panels of Figure \ref{fig:SYKspectrum}.

\subsection{ETH for the hopping operator}\label{app.hopping}
Everything we said about thermalization in eigenstates should hold for any non-extensive operator in SYK model. In order to illustrate that this is indeed the case, we collect here some illustrative results for an operator which differs considerably from the on-site number operator, namely the two-site hopping operator.

A further Hermitian operator of interest is the two-site hopping operator $h_{ij}$. It is defined by fixing two arbitrary sites $i$ and $j$ and then writing
\be
h_{ij} = c^\dagger_i c_j + c^\dagger_j c_i\,.
\ee
One might think that the simpler operator $c_i^\dagger + c_i$ would also have been a possible choice, but it is easy to see that does not conserve fermion number and so its matrix elements vanish in a fixed fermion sector as considered in this work.

\begin{figure}[h!]
\begin{center}
\includegraphics[width=0.49\textwidth]{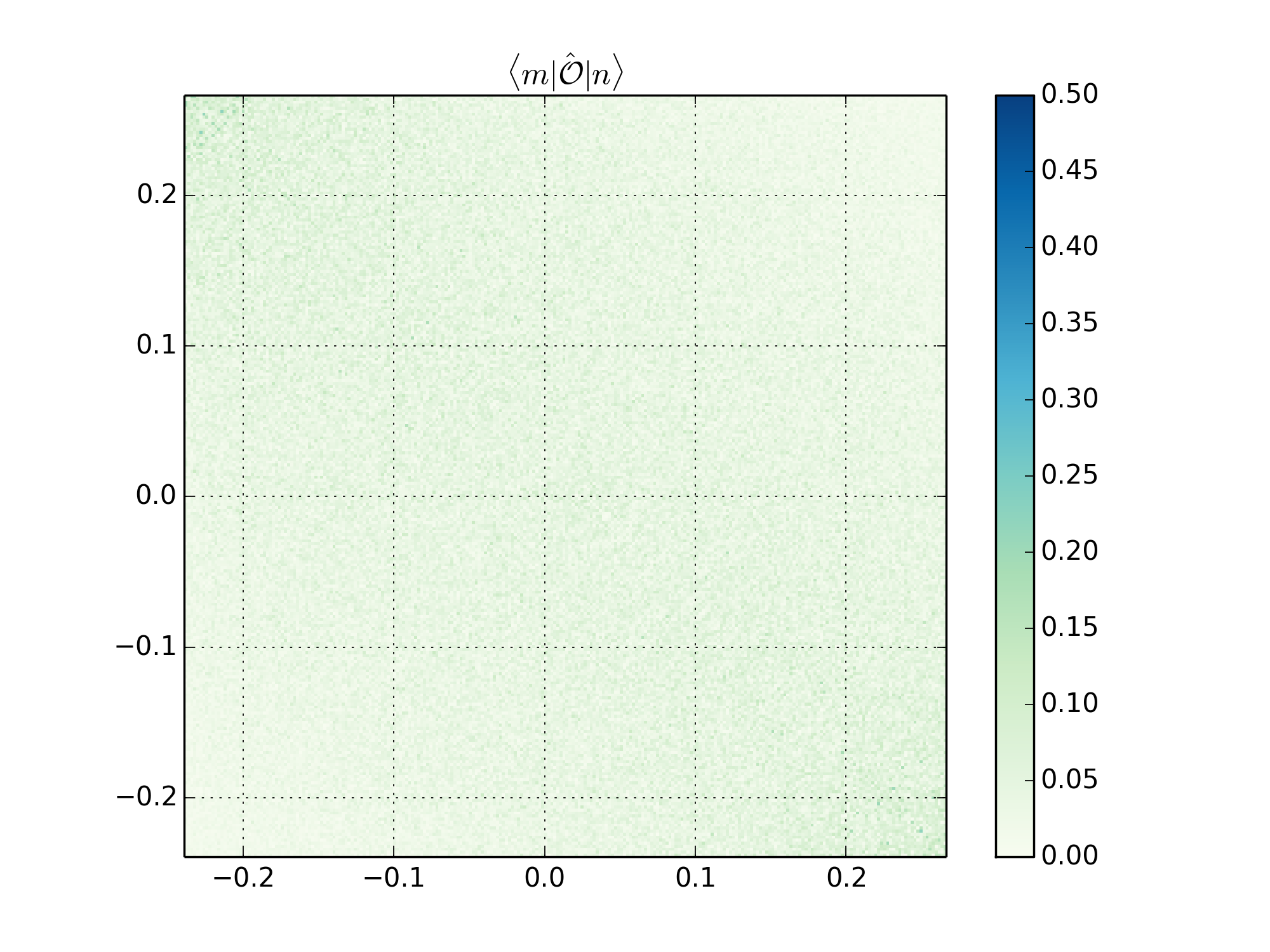} \includegraphics[width=0.49\textwidth]{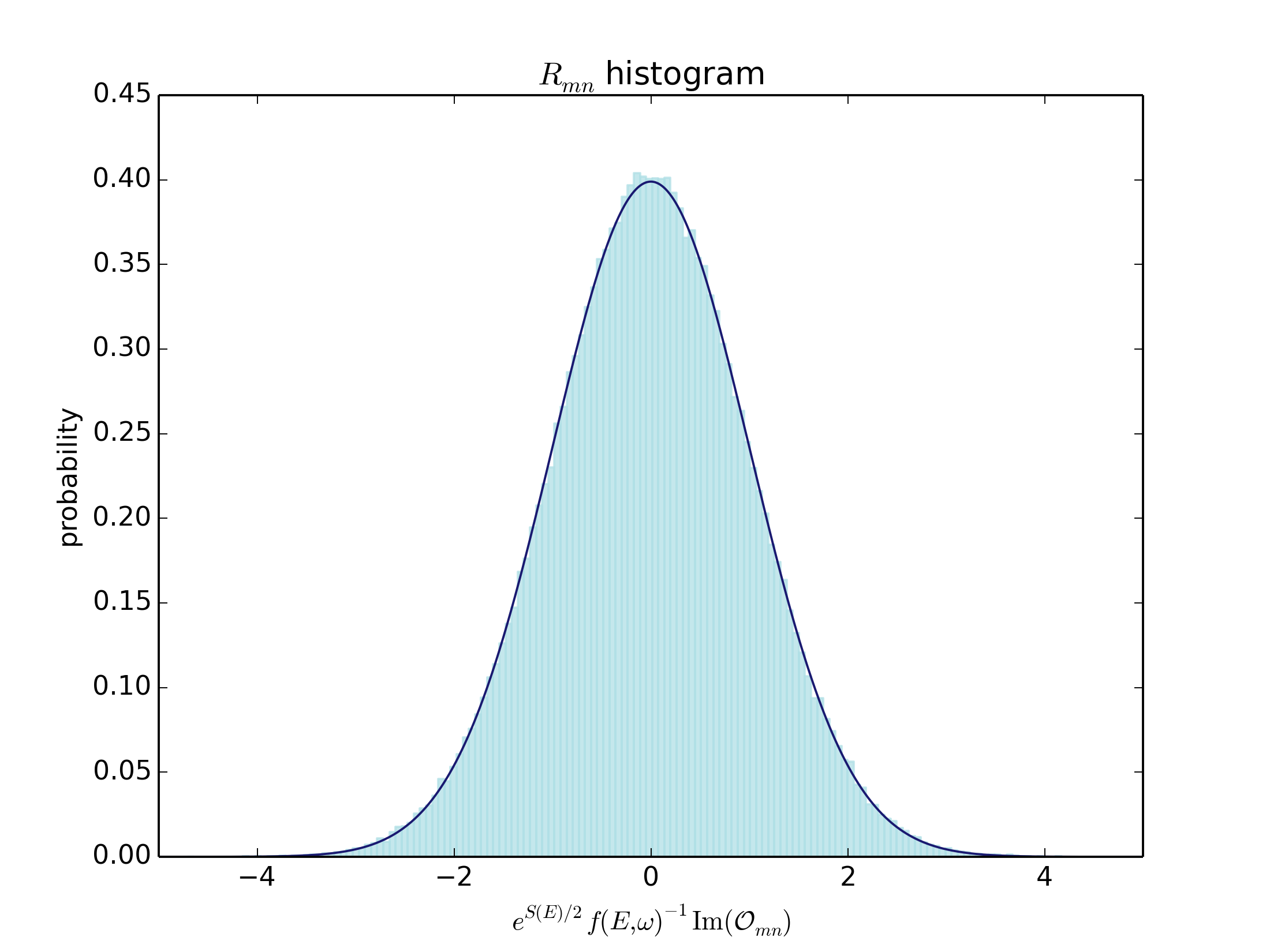}
\caption{\small Absolute values of matrix elements $|{\cal Q}_{nm}| = \left|\langle n | {\cal Q} | m \rangle\right| $  for the two-site hopping operator ${\cal Q} = \hat h_{N-1,N}$ at half filling $\nu = \frac{1}{2}$  for a single realization. {\bf Left panel}: we show the absolute values of matrix elements against their energies $E_n/J$ for $N=10$, labelled along horizontal and vertical axes. The reason one sees no clear structure along the diagonal is that the thermal expectation value $\overline{\cal Q}(\bar E) \approx 0$ for almost all energies as illustrated below in Figure \ref{fig:SYKhopdiagonal}.  {\bf Right panel: } Histogram of the remainders $R_{mn}$ for $1000$ realizations at $N=12$. As we see these are accurately fit by a unit width Gaussian with zero mean confirming the ETH ansatz (\ref{eq.ETHHypothesis}). Again we have verified this for other accessible values of $N$. \label{fig:SYKhMatrixl}}
\end{center}
\end{figure}

\begin{figure}[h!]
\begin{center}
\includegraphics[width=0.48\textwidth]{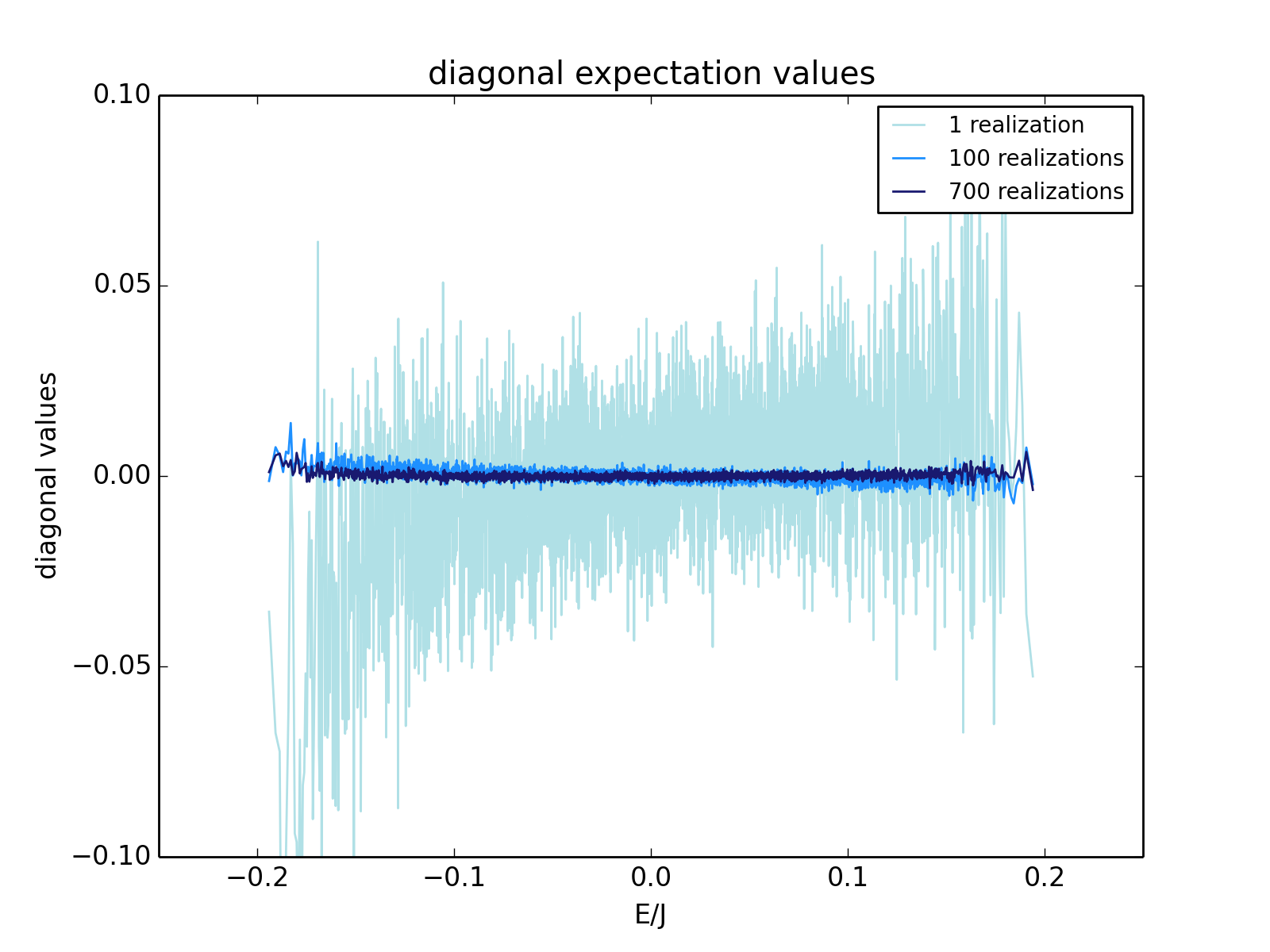}\, \includegraphics[width=0.48\textwidth]{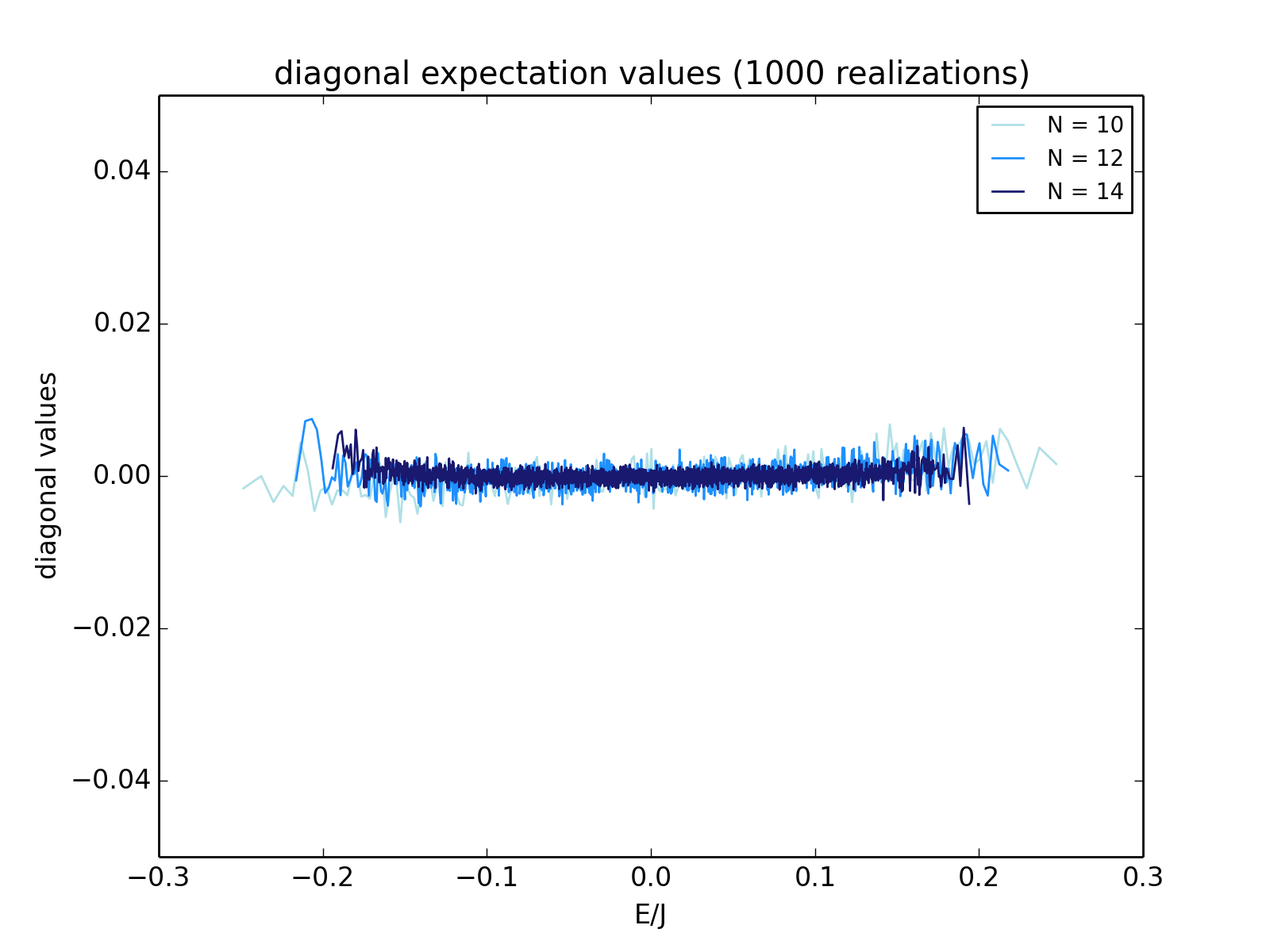}\
\caption{\small Diagonal expectation values for the two-site hopping operator $h_{N-1,N}$,  at half filling $\nu = \frac{1}{2}$.  {\bf Left panel}:   the effect of averaging of the random couplings at given fixed Hilbert space dimension, $N=14$. As expected the on-diagonal values of the ensemble approach closer and closer to a smooth curve.   {\bf Right panel}: a single random realization for increasing Hilbert space dimension corresponding to $N=12, N=14, N=16$. We see that the on-diagonal expectation values of a single realization approaches closer and closer to a smooth curve. \label{fig:SYKhopdiagonal}}
\end{center}
\end{figure}

We have extensively studied the matrix elements of the hopping operator, finding that they also satisfy the ETH ansatz. Fig \ref{fig:SYKhMatrixl} illustrates the exponential suppression of off-diagonal matrix elements by an entropy factor. The subtlety for the hopping operator is that its thermal value, i.e. its on-diagonal matrix elements, is actually zero. This is shown in Figure \ref{fig:SYKhopdiagonal}. As before we carefully distinguish between the behavior of a single randomly chosen SYK Hamiltonian (left panel) and the average over a large number of realizations (right panel).

\newpage
\section{Random matrix theory}\label{app:RMT}
In this appendix we collect some results on random matrix theory, which have been referred to occasionally in the main text. These serve as a reference for the behavior of the complex spinless SYK model, which, as we explained, shows RMT-like behavior for some parameter and energy ranges. The RMT results also served as helpful test cases to verify our algorithm with the aid of known results. 

\subsection{Off-diagonal matrix elements}
\begin{figure}[t]
\begin{center}
\includegraphics[width=0.49\textwidth]{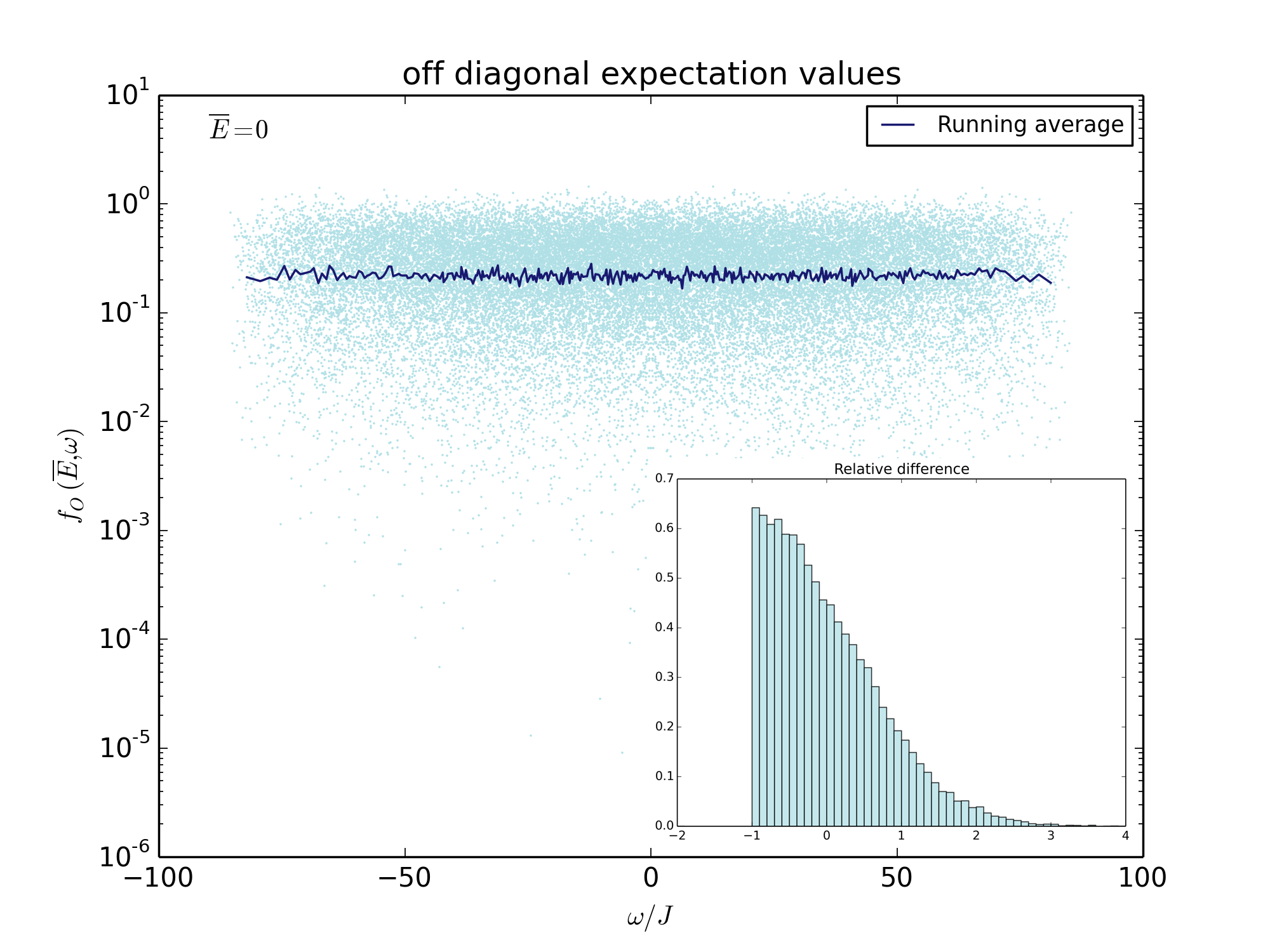} \includegraphics[width=0.49\textwidth]{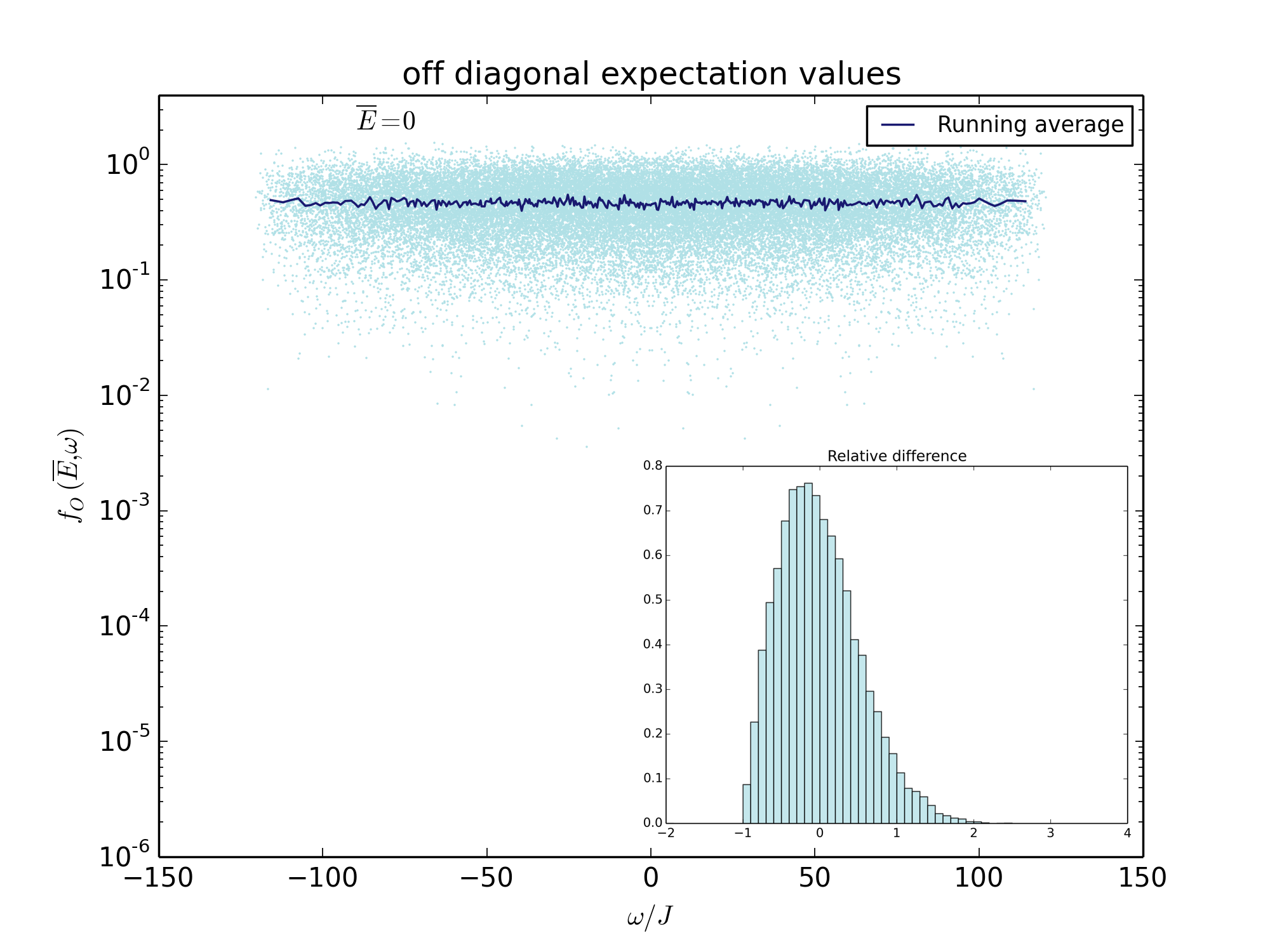}
\caption{\small Off-diagonal values of matrix elements ${\cal O}_{nm}= \left|\langle n | {\cal O} | m \rangle\right| $  for a randomly chosen operator ${\cal O}$ in random matrix theory (a GOE matrix on the left and a GUE matrix on the right) averaged over $1000$ realizations. The inset shows a histogram of relative differences between raw data and running average (within a selected energy window) {\bf Left panel: } the function $f_{\cal O}(\bar E, \omega)$ (we show the running average in dark blue and the raw data in light blue) in the GOE for Hilbert space dimension dim$({\cal H}) = 2^{12}$. One clearly sees the constancy of $f_{\cal O}(\bar E, \omega)$ in RMT. {\bf Right panel: } the function $f_{\cal O}(\bar E, \omega)$ (we show the running average in dark blue and the raw data in light blue) in the GOE for Hilbert space dimension dim$({\cal H}) = 2^{12}$. One clearly sees the constancy of $f_{\cal O}(\bar E, \omega)$ in RMT. \label{fig:RMTAppendixl}}
\end{center}
\end{figure}
Let us begin by studying the off-diagonal matrix elements
\be
{\cal O}_{mn} = \langle m | {\cal O} | n\rangle, \qquad m\neq n,
\ee
where $\left\{|n\rangle \right\}$ is the set of eigenstates of the RMT Hamiltonian and ${\cal O}$ is itself a randomly selected Hermitian operator. Of course we average all quantities over the RMT ensemble, in practice by taking the average over a large number of individual draws from the ensemble.
\subsection{Spectral form factor}
The RMT spectral form factor has recently been studied by various groups \cite{Cotler:2016fpe,delCampo:2017bzr,Cotler:2017jue} as a reference and illustration for certain aspects of the SYK case, which are qualitatively well captured in RMT. For convenience we also present this quantity here, focusing on the GUE.

\begin{figure}[t!]
\begin{center}
\includegraphics[width=0.48\textwidth]{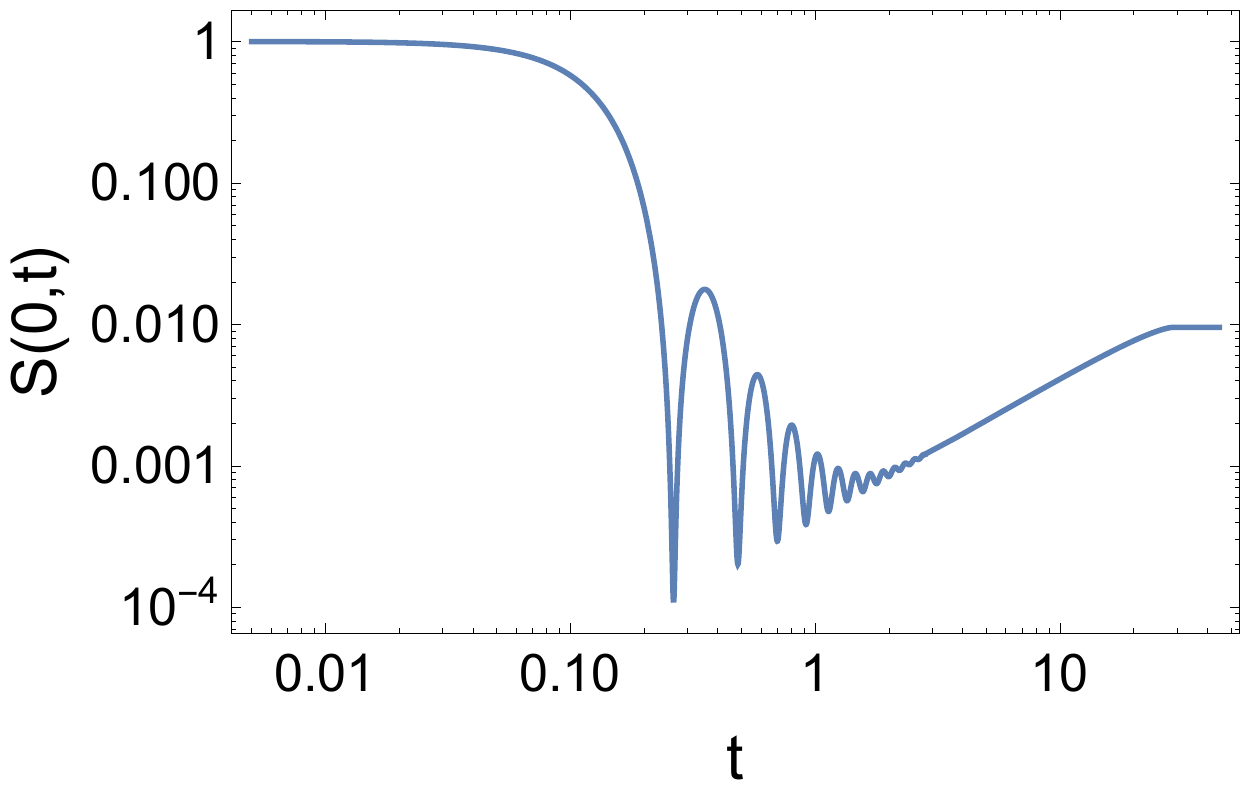}\, 
\caption{\small Spectral form factor in the GUE of size $L=105$ at $\beta =0$. The figure shows a plot of the analytic function (\ref{eq.spectralGUE}). \label{fig:SpectralGUE}}
\end{center}
\end{figure}

In RMT, more specifically the GUE, it is actually possible to analytically calculate the spectral form factor \cite{PhysRevE.55.4067,delCampo:2017bzr} using the method of orthogonal polynomials. Let $L$ be the Hilbert space dimension, that is to say we consider the ensemble of $L\times L$ Hermitian matrices. For SYK the Hilbert space dimension was $2^N$ so, when comparing the two one should obviously set $2^N = L$. If we define $\nu = \beta + i t$, the answer takes the form \cite{delCampo:2017bzr}
\be
{\cal S}(\beta,t)  = \frac{g(\beta,t)}{Z(\beta)^2}
\ee
with
\be\label{eq.spectralGUE}
g(\beta,t) = e^{\frac{1}{4}(\nu^2 + \bar\nu^2)} \left(g_c(\beta,t) + g_d(\beta,t)\right) + Z(2\beta).
\ee
The connected part is compactly expressed as
\be
g_c(\beta ,t) = \sum_{j,k}^{L-1} \left(\frac{\nu}{\bar\nu}\right)^{k-j} \left|\psi_{jk} \left( -\frac{\nu^2}{2}\right) \right|^2
\ee
with $\psi_{ij}(x)$ is given in terms of an associated Laguerre polynomial ${\sf L}_{i}^{j-i}(x)$
\be
\psi_{ij}(x) = \frac{\Gamma(i+1)}{\Gamma(j+1)}{\sf L}_{i}^{j-i}(x)
\ee
and the disconnected part as
\be
g_d(\beta ,t) = \left| {\sf L}_{L-1}^1 \left( -\frac{\nu^2}{2}\right)\right|^2
\ee
in terms of the associated Laguerre polynomial ${\sf L}_{L-1}^1(x)$. We show a plot of the function (\ref{eq.spectralGUE}) in Figure \ref{fig:SpectralGUE}. One can clearly appreciate the qualitative ressemblance to the SYK case. A detailed discussion of time scales and various power laws can be found in \cite{delCampo:2017bzr}.

\bibliographystyle{utphys}
\bibliography{SYKETHrefs}{}

\providecommand{\href}[2]{#2}\begingroup\raggedright\begin{thebibliography}{10}

\bibitem{hawking1976breakdown}
S.~W. Hawking, ``Breakdown of predictability in gravitational collapse,'' {\em
  Physical Review D} {\bfseries 14} no.~10, (1976) 2460.

\bibitem{Almheiri:2012rt}
A.~Almheiri, D.~Marolf, J.~Polchinski, and J.~Sully, ``{Black Holes:
  Complementarity Or Firewalls?},''
  \href{http://dx.doi.org/10.1007/JHEP02(2013)062}{{\em JHEP} {\bfseries 02}
  (2013) 062},
\href{http://arxiv.org/abs/1207.3123}{{\ttfamily arXiv:1207.3123 [hep-th]}}.

\bibitem{deutsch1991quantum}
J.~Deutsch, ``Quantum statistical mechanics in a closed system,'' {\em Physical
  Review A} {\bfseries 43} no.~4, (1991) 2046.

\bibitem{srednicki1994chaos}
M.~Srednicki, ``Chaos and quantum thermalization,'' {\em Physical Review E}
  {\bfseries 50} no.~2, (1994) 888.

\bibitem{Fitzpatrick:2015foa}
A.~L. Fitzpatrick, J.~Kaplan, M.~T. Walters, and J.~Wang, ``{Hawking from
  Catalan},''
\href{http://arxiv.org/abs/1510.00014}{{\ttfamily arXiv:1510.00014 [hep-th]}}.

\bibitem{Fitzpatrick:2016ive}
A.~L. Fitzpatrick, J.~Kaplan, D.~Li, and J.~Wang, ``{On information loss in
  AdS$_{3}$/CFT$_{2}$},'' \href{http://dx.doi.org/10.1007/JHEP05(2016)109}{{\em
  JHEP} {\bfseries 05} (2016) 109},
\href{http://arxiv.org/abs/1603.08925}{{\ttfamily arXiv:1603.08925 [hep-th]}}.

\bibitem{Anous:2016kss}
T.~Anous, T.~Hartman, A.~Rovai, and J.~Sonner, ``{Black Hole Collapse in the
  1/c Expansion},'' \href{http://dx.doi.org/10.1007/JHEP07(2016)123}{{\em JHEP}
  {\bfseries 07} (2016) 123},
\href{http://arxiv.org/abs/1603.04856}{{\ttfamily arXiv:1603.04856 [hep-th]}}.

\bibitem{Anous:2017tza}
T.~Anous, T.~Hartman, A.~Rovai, and J.~Sonner, ``{From Conformal Blocks to Path
  Integrals in the Vaidya Geometry},''
\href{http://arxiv.org/abs/1706.02668}{{\ttfamily arXiv:1706.02668 [hep-th]}}.

\bibitem{AK15}
A.~Kitaev, ``A simple model of quantum holography.'' Talks at KITP, April 7,
  2015 and May 27, 2015.

\bibitem{Sachdev:2015efa}
S.~Sachdev, ``{Bekenstein-Hawking Entropy and Strange Metals},''
  \href{http://dx.doi.org/10.1103/PhysRevX.5.041025}{{\em Phys. Rev.}
  {\bfseries X5} no.~4, (2015) 041025},
\href{http://arxiv.org/abs/1506.05111}{{\ttfamily arXiv:1506.05111 [hep-th]}}.

\bibitem{Maldacena:2016hyu}
J.~Maldacena and D.~Stanford, ``{Remarks on the Sachdev-Ye-Kitaev model},''
  \href{http://dx.doi.org/10.1103/PhysRevD.94.106002}{{\em Phys. Rev.}
  {\bfseries D94} no.~10, (2016) 106002},
\href{http://arxiv.org/abs/1604.07818}{{\ttfamily arXiv:1604.07818 [hep-th]}}.

\bibitem{Maldacena:2016upp}
J.~Maldacena, D.~Stanford, and Z.~Yang, ``{Conformal symmetry and its breaking
  in two dimensional Nearly Anti-de-Sitter space},''
  \href{http://dx.doi.org/10.1093/ptep/ptw124}{{\em PTEP} {\bfseries 2016}
  no.~12, (2016) 12C104},
\href{http://arxiv.org/abs/1606.01857}{{\ttfamily arXiv:1606.01857 [hep-th]}}.

\bibitem{Sachdev:1992fk}
S.~Sachdev and J.~Ye, ``{Gapless spin fluid ground state in a random, quantum
  Heisenberg magnet},''
  \href{http://dx.doi.org/10.1103/PhysRevLett.70.3339}{{\em Phys. Rev. Lett.}
  {\bfseries 70} (1993) 3339},
\href{http://arxiv.org/abs/cond-mat/9212030}{{\ttfamily arXiv:cond-mat/9212030
  [cond-mat]}}.

\bibitem{Polchinski:2016xgd}
J.~Polchinski and V.~Rosenhaus, ``{The Spectrum in the Sachdev-Ye-Kitaev
  Model},'' \href{http://dx.doi.org/10.1007/JHEP04(2016)001}{{\em JHEP}
  {\bfseries 04} (2016) 001},
\href{http://arxiv.org/abs/1601.06768}{{\ttfamily arXiv:1601.06768 [hep-th]}}.

\bibitem{Maldacena:2015waa}
J.~Maldacena, S.~H. Shenker, and D.~Stanford, ``{A bound on chaos},''
  \href{http://dx.doi.org/10.1007/JHEP08(2016)106}{{\em JHEP} {\bfseries 08}
  (2016) 106},
\href{http://arxiv.org/abs/1503.01409}{{\ttfamily arXiv:1503.01409 [hep-th]}}.

\bibitem{Bagrets:2016pxi}
D.~Bagrets, A.~Altland, and A.~Kamenev, ``{Sachdev–Ye–Kitaev model as
  Liouville quantum mechanics},''
\newblock
2016.
\newblock

\bibitem{Bagrets:2017pwq}
D.~Bagrets, A.~Altland, and A.~Kamenev, ``{Power-law out of time order
  correlation functions in the SYK model},''
  \href{http://dx.doi.org/10.1016/j.nuclphysb.2017.06.012}{{\em Nucl. Phys.}
  {\bfseries B921} (2017) 727--752},
\href{http://arxiv.org/abs/1702.08902}{{\ttfamily arXiv:1702.08902
  [cond-mat.str-el]}}.

\bibitem{Danshita:2016xbo}
I.~Danshita, M.~Hanada, and M.~Tezuka, ``{Creating and probing the
  Sachdev-Ye-Kitaev model with ultracold gases: Towards experimental studies of
  quantum gravity},''
\href{http://arxiv.org/abs/1606.02454}{{\ttfamily arXiv:1606.02454
  [cond-mat.quant-gas]}}.

\bibitem{Garcia-Alvarez:2016wem}
L.~García-Álvarez, I.~L. Egusquiza, L.~Lamata, A.~del Campo, J.~Sonner, and
  E.~Solano, ``{Digital Quantum Simulation of Minimal AdS/CFT},'' {\em Phys.
  Rev. Lett. {\it in press}} (2016) ,
\href{http://arxiv.org/abs/1607.08560}{{\ttfamily arXiv:1607.08560
  [quant-ph]}}.

\bibitem{Pikulin:2017mhj}
D.~I. Pikulin and M.~Franz, ``{Black hole on a chip: proposal for a physical
  realization of the SYK model in a solid-state system},''
  \href{http://dx.doi.org/10.1103/PhysRevX.7.031006}{{\em Phys. Rev.}
  {\bfseries X7} no.~3, (2017) 031006},
\href{http://arxiv.org/abs/1702.04426}{{\ttfamily arXiv:1702.04426
  [cond-mat.dis-nn]}}.

\bibitem{Chew:2017xuo}
A.~Chew, A.~Essin, and J.~Alicea, ``{Approximating the Sachdev-Ye-Kitaev model
  with Majorana wires},''
\href{http://arxiv.org/abs/1703.06890}{{\ttfamily arXiv:1703.06890
  [cond-mat.dis-nn]}}.

\bibitem{d2015quantum}
L.~D'Alessio, Y.~Kafri, A.~Polkovnikov, and M.~Rigol, ``From quantum chaos and
  eigenstate thermalization to statistical mechanics and thermodynamics,'' {\em
  Advances in Physics} {\bfseries 65} no.~3, (2016) 239--362.

\bibitem{Witten:2016iux}
E.~Witten, ``{An SYK-Like Model Without Disorder},''
\href{http://arxiv.org/abs/1610.09758}{{\ttfamily arXiv:1610.09758 [hep-th]}}.

\bibitem{Klebanov:2016xxf}
I.~R. Klebanov and G.~Tarnopolsky, ``{Uncolored random tensors, melon diagrams,
  and the Sachdev-Ye-Kitaev models},''
  \href{http://dx.doi.org/10.1103/PhysRevD.95.046004}{{\em Phys. Rev.}
  {\bfseries D95} no.~4, (2017) 046004},
\href{http://arxiv.org/abs/1611.08915}{{\ttfamily arXiv:1611.08915 [hep-th]}}.

\bibitem{Gurau:2016lzk}
R.~Gurau, ``{The complete $1/N$ expansion of a SYK–like tensor model},''
  \href{http://dx.doi.org/10.1016/j.nuclphysb.2017.01.015}{{\em Nucl. Phys.}
  {\bfseries B916} (2017) 386--401},
\href{http://arxiv.org/abs/1611.04032}{{\ttfamily arXiv:1611.04032 [hep-th]}}.

\bibitem{Khlebnikov:2013yia}
S.~Khlebnikov and M.~Kruczenski, ``{Thermalization of isolated quantum
  systems},''
\href{http://arxiv.org/abs/1312.4612}{{\ttfamily arXiv:1312.4612
  [cond-mat.stat-mech]}}.

\bibitem{Eberlein:2017wah}
A.~Eberlein, V.~Kasper, S.~Sachdev, and J.~Steinberg, ``{Quantum quench of the
  Sachdev-Ye-Kitaev Model},''
\href{http://arxiv.org/abs/1706.07803}{{\ttfamily arXiv:1706.07803
  [cond-mat.str-el]}}.

\bibitem{Magan:2015yoa}
J.~M. Magan, ``{Random free fermions: An analytical example of eigenstate
  thermalization},''
  \href{http://dx.doi.org/10.1103/PhysRevLett.116.030401}{{\em Phys. Rev.
  Lett.} {\bfseries 116} no.~3, (2016) 030401},
\href{http://arxiv.org/abs/1508.05339}{{\ttfamily arXiv:1508.05339
  [quant-ph]}}.

\bibitem{Kourkoulou:2017zaj}
I.~Kourkoulou and J.~Maldacena, ``{Pure states in the SYK model and
  nearly-$AdS_2$ gravity},''
\href{http://arxiv.org/abs/1707.02325}{{\ttfamily arXiv:1707.02325 [hep-th]}}.

\bibitem{Erdmenger:2016msd}
J.~Erdmenger, M.~Flory, M.-N. Newrzella, M.~Strydom, and J.~M.~S. Wu,
  ``{Quantum Quenches in a Holographic Kondo Model},''
  \href{http://dx.doi.org/10.1007/JHEP04(2017)045}{{\em JHEP} {\bfseries 04}
  (2017) 045},
\href{http://arxiv.org/abs/1612.06860}{{\ttfamily arXiv:1612.06860 [hep-th]}}.

\bibitem{Withers:2016lft}
B.~Withers, ``{Nonlinear conductivity and the ringdown of currents in metallic
  holography},'' \href{http://dx.doi.org/10.1007/JHEP10(2016)008}{{\em JHEP}
  {\bfseries 10} (2016) 008},
\href{http://arxiv.org/abs/1606.03457}{{\ttfamily arXiv:1606.03457 [hep-th]}}.

\bibitem{rigol2007thermalization}
M.~Rigol, V.~Dunjko, and M.~Olshanii, ``Thermalization and its mechanism for
  generic isolated quantum systems,''
  \href{http://dx.doi.org/10.1038/nature06838}{{\em Nature} {\bfseries 452}
  (2008) 854--858}, \href{http://arxiv.org/abs/0708.1324}{{\ttfamily
  arXiv:0708.1324 [cond-mat]}}.

\bibitem{Fu:2016yrv}
W.~Fu and S.~Sachdev, ``{Numerical study of fermion and boson models with
  infinite-range random interactions},''
  \href{http://dx.doi.org/10.1103/PhysRevB.94.035135}{{\em Phys. Rev.}
  {\bfseries B94} no.~3, (2016) 035135},
\href{http://arxiv.org/abs/1603.05246}{{\ttfamily arXiv:1603.05246
  [cond-mat.str-el]}}.

\bibitem{parcollet1999non}
O.~Parcollet and A.~Georges, ``Non-fermi-liquid regime of a doped mott
  insulator,'' \href{http://dx.doi.org/10.1103/PhysRevB.59.5341}{{\em Phys.
  Rev. B} {\bfseries 59} (Feb, 1999) 5341--5360}.
  \url{https://link.aps.org/doi/10.1103/PhysRevB.59.5341}.

\bibitem{Jevicki:2016bwu}
A.~Jevicki, K.~Suzuki, and J.~Yoon, ``{Bi-Local Holography in the SYK Model},''
  \href{http://dx.doi.org/10.1007/JHEP07(2016)007}{{\em JHEP} {\bfseries 07}
  (2016) 007},
\href{http://arxiv.org/abs/1603.06246}{{\ttfamily arXiv:1603.06246 [hep-th]}}.

\bibitem{Gross:2017hcz}
D.~J. Gross and V.~Rosenhaus, ``{The Bulk Dual of SYK: Cubic Couplings},''
  \href{http://dx.doi.org/10.1007/JHEP05(2017)092}{{\em JHEP} {\bfseries 05}
  (2017) 092},
\href{http://arxiv.org/abs/1702.08016}{{\ttfamily arXiv:1702.08016 [hep-th]}}.

\bibitem{Jevicki:2016ito}
A.~Jevicki and K.~Suzuki, ``{Bi-Local Holography in the SYK Model:
  Perturbations},'' \href{http://dx.doi.org/10.1007/JHEP11(2016)046}{{\em JHEP}
  {\bfseries 11} (2016) 046},
\href{http://arxiv.org/abs/1608.07567}{{\ttfamily arXiv:1608.07567 [hep-th]}}.

\bibitem{Dartois:2017xoe}
S.~Dartois, H.~Erbin, and S.~Mondal, ``{Conformality of $1/N$ corrections in
  SYK-like models},''
\href{http://arxiv.org/abs/1706.00412}{{\ttfamily arXiv:1706.00412 [hep-th]}}.

\bibitem{Shenker:2013pqa}
S.~H. Shenker and D.~Stanford, ``{Black holes and the butterfly effect},''
  \href{http://dx.doi.org/10.1007/JHEP03(2014)067}{{\em JHEP} {\bfseries 03}
  (2014) 067},
\href{http://arxiv.org/abs/1306.0622}{{\ttfamily arXiv:1306.0622 [hep-th]}}.

\bibitem{rigol2009quantum}
M.~Rigol, ``Quantum quenches and thermalization in one-dimensional fermionic
  systems,'' \href{http://dx.doi.org/10.1103/PhysRevA.80.053607}{{\em Phys.
  Rev. A} {\bfseries 80} (Nov, 2009) 053607}.
  \url{https://link.aps.org/doi/10.1103/PhysRevA.80.053607}.

\bibitem{Ferrari:2017ryl}
F.~Ferrari, ``{The Large D Limit of Planar Diagrams},''
\href{http://arxiv.org/abs/1701.01171}{{\ttfamily arXiv:1701.01171 [hep-th]}}.

\bibitem{beugeling2015off}
W.~Beugeling, R.~Moessner, and M.~Haque, ``Off-diagonal matrix elements of
  local operators in many-body quantum systems,'' {\em Physical Review E}
  {\bfseries 91} no.~1, (2015) 012144.

\bibitem{Garcia-Garcia:2016mno}
A.~M. García-García and J.~J.~M. Verbaarschot, ``{Spectral and thermodynamic
  properties of the Sachdev-Ye-Kitaev model},''
  \href{http://dx.doi.org/10.1103/PhysRevD.94.126010}{{\em Phys. Rev.}
  {\bfseries D94} no.~12, (2016) 126010},
\href{http://arxiv.org/abs/1610.03816}{{\ttfamily arXiv:1610.03816 [hep-th]}}.

\bibitem{Cotler:2016fpe}
J.~S. Cotler, G.~Gur-Ari, M.~Hanada, J.~Polchinski, P.~Saad, S.~H. Shenker,
  D.~Stanford, A.~Streicher, and M.~Tezuka, ``{Black Holes and Random
  Matrices},'' \href{http://dx.doi.org/10.1007/JHEP05(2017)118}{{\em JHEP}
  {\bfseries 05} (2017) 118},
\href{http://arxiv.org/abs/1611.04650}{{\ttfamily arXiv:1611.04650 [hep-th]}}.

\bibitem{Davison:2016ngz}
R.~A. Davison, W.~Fu, A.~Georges, Y.~Gu, K.~Jensen, and S.~Sachdev,
  ``{Thermoelectric transport in disordered metals without quasiparticles: The
  Sachdev-Ye-Kitaev models and holography},''
  \href{http://dx.doi.org/10.1103/PhysRevB.95.155131}{{\em Phys. Rev.}
  {\bfseries B95} no.~15, (2017) 155131},
\href{http://arxiv.org/abs/1612.00849}{{\ttfamily arXiv:1612.00849
  [cond-mat.str-el]}}.

\bibitem{delCampo:2017bzr}
A.~del Campo, J.~Molina-Vilaplana, and J.~Sonner, ``{Scrambling the spectral
  form factor: unitarity constraints and exact results},''
  \href{http://dx.doi.org/10.1103/PhysRevD.95.126008}{{\em Phys. Rev.}
  {\bfseries D95} no.~12, (2017) 126008},
\href{http://arxiv.org/abs/1702.04350}{{\ttfamily arXiv:1702.04350 [hep-th]}}.

\bibitem{PhysRevE.55.4067}
E.~Br\'ezin and S.~Hikami, ``Spectral form factor in a random matrix theory,''
  \href{http://dx.doi.org/10.1103/PhysRevE.55.4067}{{\em Phys. Rev. E}
  {\bfseries 55} (Apr, 1997) 4067--4083}.
  \url{https://link.aps.org/doi/10.1103/PhysRevE.55.4067}.

\bibitem{tavora2016inevitable}
M.~T{\'a}vora, E.~Torres-Herrera, and L.~F. Santos, ``Inevitable power-law
  behavior of isolated many-body quantum systems and how it anticipates
  thermalization,'' {\em Physical Review A} {\bfseries 94} no.~4, (2016)
  041603.

\bibitem{Krishnan:2017txw}
C.~Krishnan and K.~V.~P. Kumar, ``{Towards a Finite-$N$ Hologram},''
\href{http://arxiv.org/abs/1706.05364}{{\ttfamily arXiv:1706.05364 [hep-th]}}.

\bibitem{Jackiw:1984je}
R.~Jackiw, ``{Lower Dimensional Gravity},''
\href{http://dx.doi.org/10.1016/0550-3213(85)90448-1}{{\em Nucl. Phys.}
  {\bfseries B252} (1985) 343--356}.

\bibitem{Teitelboim:1983ux}
C.~Teitelboim, ``{Gravitation and Hamiltonian Structure in Two Space-Time
  Dimensions},''
\href{http://dx.doi.org/10.1016/0370-2693(83)90012-6}{{\em Phys. Lett.}
  {\bfseries 126B} (1983) 41--45}.

\bibitem{Almheiri:2014cka}
A.~Almheiri and J.~Polchinski, ``{Models of AdS$_{2}$ backreaction and
  holography},'' \href{http://dx.doi.org/10.1007/JHEP11(2015)014}{{\em JHEP}
  {\bfseries 11} (2015) 014},
\href{http://arxiv.org/abs/1402.6334}{{\ttfamily arXiv:1402.6334 [hep-th]}}.

\bibitem{Jensen:2016pah}
K.~Jensen, ``{Chaos in AdS$_2$ Holography},''
  \href{http://dx.doi.org/10.1103/PhysRevLett.117.111601}{{\em Phys. Rev.
  Lett.} {\bfseries 117} no.~11, (2016) 111601},
\href{http://arxiv.org/abs/1605.06098}{{\ttfamily arXiv:1605.06098 [hep-th]}}.

\bibitem{Mandal:2017thl}
G.~Mandal, P.~Nayak, and S.~R. Wadia, ``{Coadjoint orbit action of Virasoro
  group and two-dimensional quantum gravity dual to SYK/tensor models},''
\href{http://arxiv.org/abs/1702.04266}{{\ttfamily arXiv:1702.04266 [hep-th]}}.

\bibitem{Taylor:2017dly}
M.~Taylor, ``{Generalized conformal structure, dilaton gravity and SYK},''
\href{http://arxiv.org/abs/1706.07812}{{\ttfamily arXiv:1706.07812 [hep-th]}}.

\bibitem{Das:2017pif}
S.~R. Das, A.~Jevicki, and K.~Suzuki, ``{Three Dimensional View of the SYK/AdS
  Duality},'' \href{http://dx.doi.org/10.1007/JHEP09(2017)017}{{\em JHEP}
  {\bfseries 09} (2017) 017},
\href{http://arxiv.org/abs/1704.07208}{{\ttfamily arXiv:1704.07208 [hep-th]}}.

\bibitem{Papadodimas:2013wnh}
K.~Papadodimas and S.~Raju, ``{Black Hole Interior in the Holographic
  Correspondence and the Information Paradox},''
  \href{http://dx.doi.org/10.1103/PhysRevLett.112.051301}{{\em Phys. Rev.
  Lett.} {\bfseries 112} no.~5, (2014) 051301},
\href{http://arxiv.org/abs/1310.6334}{{\ttfamily arXiv:1310.6334 [hep-th]}}.

\bibitem{Maldacena:2001kr}
J.~M. Maldacena, ``{Eternal Black Holes in Anti-de~Sitter},''
  \href{http://dx.doi.org/10.1088/1126-6708/2003/04/021}{{\em JHEP} {\bfseries
  04} (2003) 021},
\href{http://arxiv.org/abs/hep-th/0106112}{{\ttfamily arXiv:hep-th/0106112
  [hep-th]}}.

\bibitem{Maldacena:2013xja}
J.~Maldacena and L.~Susskind, ``{Cool horizons for entangled black holes},''
  \href{http://dx.doi.org/10.1002/prop.201300020}{{\em Fortsch. Phys.}
  {\bfseries 61} (2013) 781--811},
\href{http://arxiv.org/abs/1306.0533}{{\ttfamily arXiv:1306.0533 [hep-th]}}.

\bibitem{Jensen:2013ora}
K.~Jensen and A.~Karch, ``{Holographic Dual of an Einstein-Podolsky-Rosen Pair
  has a Wormhole},''
  \href{http://dx.doi.org/10.1103/PhysRevLett.111.211602}{{\em Phys. Rev.
  Lett.} {\bfseries 111} no.~21, (2013) 211602},
\href{http://arxiv.org/abs/1307.1132}{{\ttfamily arXiv:1307.1132 [hep-th]}}.

\bibitem{Sonner:2013mba}
J.~Sonner, ``{Holographic Schwinger Effect and the Geometry of Entanglement},''
  \href{http://dx.doi.org/10.1103/PhysRevLett.111.211603}{{\em Phys. Rev.
  Lett.} {\bfseries 111} no.~21, (2013) 211603},
\href{http://arxiv.org/abs/1307.6850}{{\ttfamily arXiv:1307.6850 [hep-th]}}.

\bibitem{Marolf:2013dba}
D.~Marolf and J.~Polchinski, ``{Gauge/Gravity Duality and the Black Hole
  Interior},'' \href{http://dx.doi.org/10.1103/PhysRevLett.111.171301}{{\em
  Phys. Rev. Lett.} {\bfseries 111} (2013) 171301},
\href{http://arxiv.org/abs/1307.4706}{{\ttfamily arXiv:1307.4706 [hep-th]}}.

\bibitem{Balasubramanian:2014gla}
V.~Balasubramanian, M.~Berkooz, S.~F. Ross, and J.~Simon, ``{Black Holes,
  Entanglement and Random Matrices},''
  \href{http://dx.doi.org/10.1088/0264-9381/31/18/185009}{{\em Class. Quant.
  Grav.} {\bfseries 31} (2014) 185009},
\href{http://arxiv.org/abs/1404.6198}{{\ttfamily arXiv:1404.6198 [hep-th]}}.

\bibitem{Papadodimas:2012aq}
K.~Papadodimas and S.~Raju, ``{An Infalling Observer in AdS/CFT},''
  \href{http://dx.doi.org/10.1007/JHEP10(2013)212}{{\em JHEP} {\bfseries 10}
  (2013) 212},
\href{http://arxiv.org/abs/1211.6767}{{\ttfamily arXiv:1211.6767 [hep-th]}}.

\bibitem{Fitzpatrick:2015zha}
A.~L. Fitzpatrick, J.~Kaplan, and M.~T. Walters, ``{Virasoro Conformal Blocks
  and Thermality from Classical Background Fields},''
  \href{http://dx.doi.org/10.1007/JHEP11(2015)200}{{\em JHEP} {\bfseries 11}
  (2015) 200},
\href{http://arxiv.org/abs/1501.05315}{{\ttfamily arXiv:1501.05315 [hep-th]}}.

\bibitem{Lashkari:2016vgj}
N.~Lashkari, A.~Dymarsky, and H.~Liu, ``{Eigenstate Thermalization Hypothesis
  in Conformal Field Theory},''
\href{http://arxiv.org/abs/1610.00302}{{\ttfamily arXiv:1610.00302 [hep-th]}}.

\bibitem{Basu:2017kzo}
P.~Basu, D.~Das, S.~Datta, and S.~Pal, ``{On thermality of CFT eigenstates},''
\href{http://arxiv.org/abs/1705.03001}{{\ttfamily arXiv:1705.03001 [hep-th]}}.

\bibitem{Anninos:2016szt}
D.~Anninos, T.~Anous, and F.~Denef, ``{Disordered Quivers and Cold Horizons},''
  \href{http://dx.doi.org/10.1007/JHEP12(2016)071}{{\em JHEP} {\bfseries 12}
  (2016) 071},
\href{http://arxiv.org/abs/1603.00453}{{\ttfamily arXiv:1603.00453 [hep-th]}}.

\bibitem{Fu:2016vas}
W.~Fu, D.~Gaiotto, J.~Maldacena, and S.~Sachdev, ``{Supersymmetric
  Sachdev-Ye-Kitaev models},''
  \href{http://dx.doi.org/10.1103/PhysRevD.95.069904,
  10.1103/PhysRevD.95.026009}{{\em Phys. Rev.} {\bfseries D95} no.~2, (2017)
  026009}, \href{http://arxiv.org/abs/1610.08917}{{\ttfamily arXiv:1610.08917
  [hep-th]}}.
[Addendum: Phys. Rev.D95,no.6,069904(2017)].

\bibitem{Li:2017hdt}
T.~Li, J.~Liu, Y.~Xin, and Y.~Zhou, ``{Supersymmetric SYK model and random
  matrix theory},''
\href{http://arxiv.org/abs/1702.01738}{{\ttfamily arXiv:1702.01738 [hep-th]}}.

\bibitem{Berkooz:2016cvq}
M.~Berkooz, P.~Narayan, M.~Rozali, and J.~Simón, ``{Higher Dimensional
  Generalizations of the SYK Model},''
  \href{http://dx.doi.org/10.1007/JHEP01(2017)138}{{\em JHEP} {\bfseries 01}
  (2017) 138},
\href{http://arxiv.org/abs/1610.02422}{{\ttfamily arXiv:1610.02422 [hep-th]}}.

\bibitem{Peng:2016mxj}
C.~Peng, M.~Spradlin, and A.~Volovich, ``{A Supersymmetric SYK-like Tensor
  Model},'' \href{http://dx.doi.org/10.1007/JHEP05(2017)062}{{\em JHEP}
  {\bfseries 05} (2017) 062},
\href{http://arxiv.org/abs/1612.03851}{{\ttfamily arXiv:1612.03851 [hep-th]}}.

\bibitem{Krishnan:2016bvg}
C.~Krishnan, S.~Sanyal, and P.~N. Bala~Subramanian, ``{Quantum Chaos and
  Holographic Tensor Models},''
  \href{http://dx.doi.org/10.1007/JHEP03(2017)056}{{\em JHEP} {\bfseries 03}
  (2017) 056},
\href{http://arxiv.org/abs/1612.06330}{{\ttfamily arXiv:1612.06330 [hep-th]}}.

\bibitem{Garcia-Garcia:2017pzl}
A.~M. García-García and J.~J.~M. Verbaarschot, ``{Analytical Spectral Density
  of the Sachdev-Ye-Kitaev Model at finite N},''
\href{http://arxiv.org/abs/1701.06593}{{\ttfamily arXiv:1701.06593 [hep-th]}}.

\bibitem{Cotler:2017jue}
J.~Cotler, N.~Hunter-Jones, J.~Liu, and B.~Yoshida, ``{Chaos, Complexity, and
  Random Matrices},''
\href{http://arxiv.org/abs/1706.05400}{{\ttfamily arXiv:1706.05400 [hep-th]}}.

\end{thebibliography}\endgroup

\end{spacing}

\end{document}